\documentclass{article}

\usepackage{microtype}
\usepackage{graphicx}
\usepackage{subcaption}
\usepackage{booktabs} % for professional tables

\usepackage{hyperref}

\usepackage[accepted]{mlsys2025}
\usepackage{graphicx}

\usepackage{subcaption}
\usepackage{caption}
\usepackage[utf8]{inputenc}
\usepackage{amsmath}
\usepackage[normalem]{ulem}
\usepackage{xcolor}
\usepackage{multirow}
\usepackage{booktabs}
\usepackage{enumitem}
\setlist[itemize]{noitemsep, nolistsep}

\newcommand{\mypar}[1]{{\smallskip \noindent \bf #1}\hspace{0.1in}}

\newcommand{\pjn}[1]{\textsc{DiffServe}}

\newcommand{\highlight}[1]{{\color{blue} #1}}

\mlsystitlerunning{\pjn{}}

\usepackage{graphicx}
\usepackage{hyperref}
\usepackage{eso-pic}
\usepackage{xparse}
\newlength{\badgewidth}
\newlength{\badgegap}
\newcommand{\badgeList}{}
\NewDocumentCommand{\addTopRightBadge}{O{} m}{%
\gappto{\badgeList}{\href{#1}{\includegraphics[width=\badgewidth]{#2}}\hspace{\badgegap}}%
}
\newcommand{\placeTopRightBadges}{%
\AddToShipoutPictureBG*{%
\put(\LenToUnit{\paperwidth - 1.5cm - \badgewidth},\LenToUnit{\paperheight - 2cm}){%
\makebox[0pt][r]{\badgeList}%
}%
}%
}
\addTopRightBadge{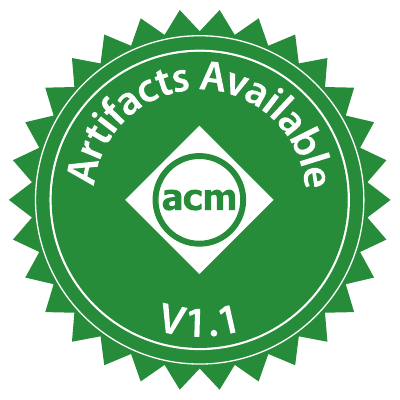}
\addTopRightBadge{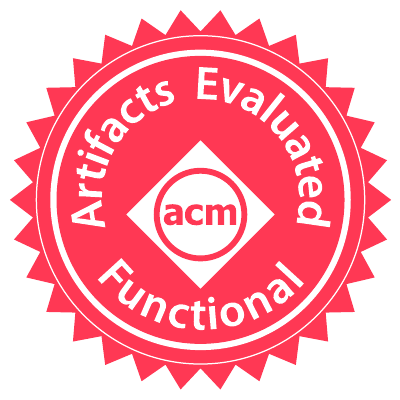}
\placeTopRightBadges

\begin{document}

\twocolumn[
\mlsystitle{DiffServe: Efficiently Serving Text-to-Image Diffusion Models with Query-Aware Model Scaling}

% unless you've provided the [accepted] option to the mlsys2024


\mlsyssetsymbol{equal}{*}

\begin{mlsysauthorlist}
\mlsysauthor{Sohaib Ahmad}{equal,umass}
\mlsysauthor{Qizheng Yang}{equal,umass}
\mlsysauthor{Haoliang Wang}{adb}
\mlsysauthor{Ramesh K. Sitaraman}{umass}
\mlsysauthor{Hui Guan}{umass}
\end{mlsysauthorlist}

\mlsysaffiliation{umass}{University of Massachusetts Amherst, Amherst, MA, USA}
\mlsysaffiliation{adb}{Adobe Research, San Jose, CA, USA}

\mlsyscorrespondingauthor{Qizheng Yang}{qizhengyang@cs.umass.edu}
\mlsyscorrespondingauthor{Sohaib Ahmad}{sohaib@cs.umass.edu}

\mlsyskeywords{Machine Learning, Systems for machine learning, Generative artificial intelligence, Diffusion, Image generation, Query aware, Scheduling, Resource management}

\vskip 0.3in
% \vskip 0.1in

\begin{abstract}
Text-to-image generation using diffusion models has gained increasing popularity due to their ability to produce high-quality, realistic images based on text prompts. However, efficiently serving these models is challenging due to their computation-intensive nature and the variation in query demands. In this paper, we aim to address both problems simultaneously through query-aware model scaling. The core idea is to construct model cascades so that easy queries can be processed by more lightweight diffusion models without compromising image generation quality. 
Based on this concept, we develop an end-to-end text-to-image diffusion model serving system, \textsc{DiffServe}, which automatically constructs model cascades from available diffusion model variants and allocates resources dynamically in response to demand fluctuations.
Our empirical evaluations demonstrate that \textsc{DiffServe} achieves up to 24\% improvement in response quality while maintaining 19-70\% lower latency violation rates compared to state-of-the-art model serving systems.

% This leads to resource savings from easy queries, thereby improving overall system throughput.
% Moreover, the serving system can accommodate fluctuating query demands by dynamically adjusting the proportion of queries treated as easy queries without a significant loss in quality. 
% This document provides a basic paper template and submission guidelines.
% Abstracts must be a single paragraph, ideally between 4--6 sentences long.
% Gross violations will trigger corrections at the camera-ready phase.
\end{abstract}
]


\printAffiliationsAndNotice{\mlsysEqualContribution} % otherwise use the standard text.

\section{Introduction}

% \TODO{Use quality instead of accuracy at all places}

% \TODO{pages: 10 pages:
% intro + motivation: 2.5 pages; 
% design: 3 pages; 
% eval: 3-4;  
% discussion, related work, conclusion: 1 page}

% what is the problem and why it is important 
% Diffusion models offer a robust framework for generating high-quality, realistic data, driving progress in areas like image synthesis and scientific simulations by adeptly modeling complex, high-dimensional distributions. In particular, 
% Text-to-image diffusion models take a text prompt as input and generate images that capture the description in the text prompt. 

Text-to-image diffusion models are a powerful class of generative models that create images from textual descriptions (queries) by progressively denoising an initial noise input into a coherent image. 
They have gained increasing popularity and been integrated into various interactive content creation workflows such as Adobe FireFly~\cite{adobefirefly} and Midjourney~\cite{midjourney}. 
With their growing adoption, it is essential to develop efficient diffusion model serving systems that deliver fast and accurate responses to queries. 
% These systems enable seamless access to pre-trained diffusion models via APIs, allowing applications to leverage generative capabilities without managing execution or scaling complexities. 
In such systems, providers must guarantee service level objectives (SLOs) to users with regard to latency deadlines. At the same time, providers aim to maximize hardware utilization by increasing throughput, i.e., serving as many queries per unit time. 

Efficiently serving text-to-image diffusion models, however, presents two main challenges. 
First, high-quality diffusion models are computationally intensive, limiting the serving throughput of the model serving system. Here, \textit{serving throughput} refers to the number of queries (i.e., text prompts) that can be processed by the system per unit of time (queries per second, QPS). 
For instance, the Stable Diffusion XL model~\cite{podell2023sdxl} achieves a 30\% improvement in generated image quality on DiffusionDB~\cite{wangDiffusionDBLargescalePrompt2022} dataset compared to the baseline SDXL-Lightning~\cite{lin2024sdxllightningprogressiveadversarialdiffusion} but is $4.6\times$ slower when processing a batch of 16 queries on a Nvidia A100-{80}GB GPU. 
This example highlights the typical trade-off between accuracy and efficiency often encountered in machine learning models.
Second, the query demand for a model serving system fluctuates over time~\cite{shahrad2020serverless, twitter_streaming_trace}. Hardware resources provisioned to handle peak demand may remain largely idle during periods of low query demand, resulting in inefficient resource utilization. 
% \TODO{This is true for Adobe but might not be for others. Might be worthwhile here or in later section to briefly mention rationale/the cost aspect of using cloud resource - on-demand resource cost a lot more than reserved resource / instance saving plans and it takes time to increase/decrease capacity, so companies tend to maintain a relatively fixed capacity for an overall lower cost}\TODO{Ramesh: Most production systems are provisioned to serve the anticipated peak load, hence have low utilization at other times.}

% \begin{figure}[t]
%   \centering
%     \includegraphics[width=0.9\linewidth]{figures/cascade.pdf}
%   \caption{The basic idea of query-aware model scaling for serving text-to-image diffusion models. D: the discriminator to identify easy queries. \TODO{Qizheng: replace the text prompt and the generated image to real prompt and generated image. Might remove the figure if space is limited.}}
%   \label{fig:baseline_comparison}
% \end{figure}

This work introduces \textit{query-aware model scaling} to tackle the challenges.
The central concept of query-aware model scaling is the creation of a \textit{diffusion model cascade}, where each query is first processed by a lightweight diffusion model (also called light model) to generate an image. 
If the output meets predefined quality requirements, as determined by a \textit{discriminator}, it is used as the final response. If not, the query is routed to a more computationally intensive but higher-quality diffusion model (also called heavy model) to produce the final result. 
The rationale is that certain queries, known as \textit{easy queries}, are inherently simpler and can be processed by small models with no or minor quality degradation.  
The approach is called \textit{query-aware} as it routes queries based on the complexities of each query.

Model cascades address the computational burden of diffusion models by allowing easy queries to be processed exclusively by light models that execute faster. 
This time saving from handling easy queries enables the system to achieve higher serving throughput. In addition, our diffusion model cascade uses a \textit{confidence threshold} to control the proportion of queries classified as easy. By adjusting this threshold, the model cascade offers a way to balance image generation quality with serving throughput. 
Leveraging the quality-throughput trade-off in this fashion to manage query demand variations is called \textit{model scaling}. Model scaling allows the system to adapt to demand variation and improves hardware utilization by avoiding the need to provision resources for peak demand.

The core idea of query-aware model scaling presents two major research challenges. 
The first challenge is to develop a discriminator that can \textit{automatically}, \textit{efficiently}, and \textit{accurately} identify easy queries. The discriminator must assess whether a generated image meets quality standards without requiring manual intervention. 
It also needs to operate efficiently to minimize runtime overhead. Moreover, accuracy in classifying the queries is essential, as routing easy queries to the heavyweight model wastes computational resources, while handling complex queries solely with the lightweight model can degrade the quality of responses. 
% While quantitative metrics like Fréchet Inception Distance (FID)~\cite{heusel2017gans} are commonly used to evaluate diffusion models, they compare how similar two datasets of images are (e.g., ground truth versus generated images) and thus cannot assess the quality of individual generated images. 
% Existing quantitative metrics, such as CLIP score~\cite{} and PickScore~\cite{}, can compute a single value for individual generated images, but 
We later show that diffusion model cascades relying on existing quantitative metrics perform no better than random classification due to their inherent limitations in capturing the nuanced image quality differences.  
The second challenge lies in efficiently allocating resources to optimize performance when serving diffusion model cascades. The key parameter in such cascades, the confidence threshold, must be co-optimized with other system parameters to achieve optimal performance. 
% These configuration parameters include the allocation ratio of hardware resources across model variants and the batch size used by each model to process queries. Properly configuring these parameters is crucial for achieving high serving throughput, high response quality, and minimizing SLO violation rates.

This work introduces \pjn{}, a system that leverages query-aware model scaling to efficiently serve text-to-image diffusion models. 
To address the first challenge, \pjn{} constructs a diffusion model cascade by training a machine learning (ML) model to assess the quality of generated images. 
The key insight is that an ML model can be trained to accurately distinguish between images generated by diffusion models and real images. 
This ML model can be repurposed to differentiate whether the images produced by the lightweight model meet quality requirements based on its classification confidence. 
To address the second challenge, \pjn{} carefully models system performance as functions of key configuration parameters and formulates the resource allocation problem in a mixed integer linear programming (MILP) framework to identify the optimal allocation plan that maximizes response quality while satisfying query demand. 
\pjn{} periodically solves it and re-allocates resources to adapt to varying query demands.

We evaluate \pjn{} on three light-heavy diffusion model pairs using both synthetic and real traces and find that it consistently outperforms state-of-the-art systems. 
Compared to Proteus~\cite{ahmad2024proteus} which leverages model scaling but randomly routes queries to model variants based on system load, \pjn{} can improve system accuracy by up to 20\% since its query routing considers their complexities. 
Compared to serving systems that use a fixed confidence threshold, \pjn{} demonstrates up to 24\% improvement in quality and 19-70\% reduction in SLO violations, owing to its better resource allocation.

We summarize the contributions of this work as follows: 
\begin{itemize}[noitemsep, nolistsep]
    \item We introduce query-aware model scaling which constructs diffusion model cascades to optimize the efficiency of diffusion model serving systems.
    \item We leverage adversarial training to build discriminators that enable the cascading of diffusion model variants.
    \item We formulate the resource allocation problem as a mixed integer linear programming (MILP) framework to determine optimal configuration parameters when serving diffusion model cascades.
    \item We implement these techniques in the \pjn{}  model serving system and evaluate its performance across various workload traces and diffusion models.
\end{itemize}

% \vspace{-1em}
\section{Motivation and Challenges}
\label{sec:motivation}
This section motivates query-aware model scaling for serving diffusion models and explains the research challenges.

% This is redudant to intro 
% Text-to-image diffusion models are a powerful class of generative models that create images from textual descriptions by progressively denoising an initial noise input into a coherent image. Models like DALL-E, GLIDE, and more recent advancements have demonstrated an exceptional ability to generate diverse, high-quality images based on text prompts. Their applications span various interactive domains, including creative content generation, design prototyping, gaming, and virtual worlds, where producing high-quality outputs under strict latency constraints is crucial.

% To address these demands, diffusion model serving systems provide high-level APIs that simplify integration, enabling users to submit queries and receive generated images within defined latency deadlines, without the need to manage the underlying execution or scaling complexities. These systems ensure both high-quality image outputs and high throughput, while meeting Service Level Objective (SLO) requirements, particularly regarding latency, making them ideal for real-time and resource-intensive applications.  

% Text-to-image diffusion models are a powerful class of generative models that create images from textual descriptions by progressively denoising an initial noise input into a coherent image. 

\begin{figure*}[t]%
    \centering
    \subfloat[\centering FID vs. average inference latency \label{fig:model-variants}]{{\includegraphics[width=.3\textwidth]{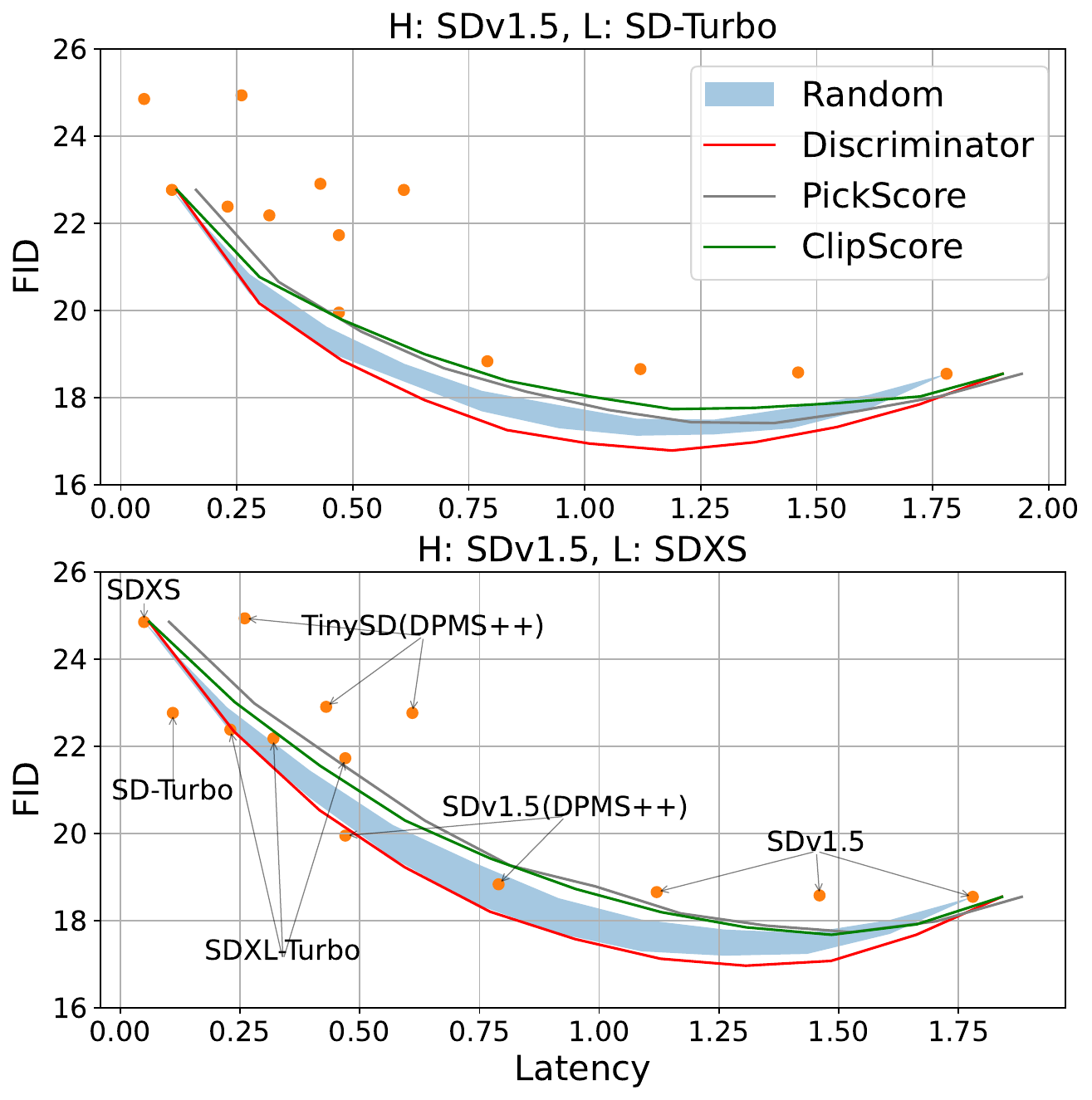}}}%
    \subfloat[\centering CDF of image quality difference \label{fig:easy-hard}]{{\includegraphics[width=.35\textwidth]{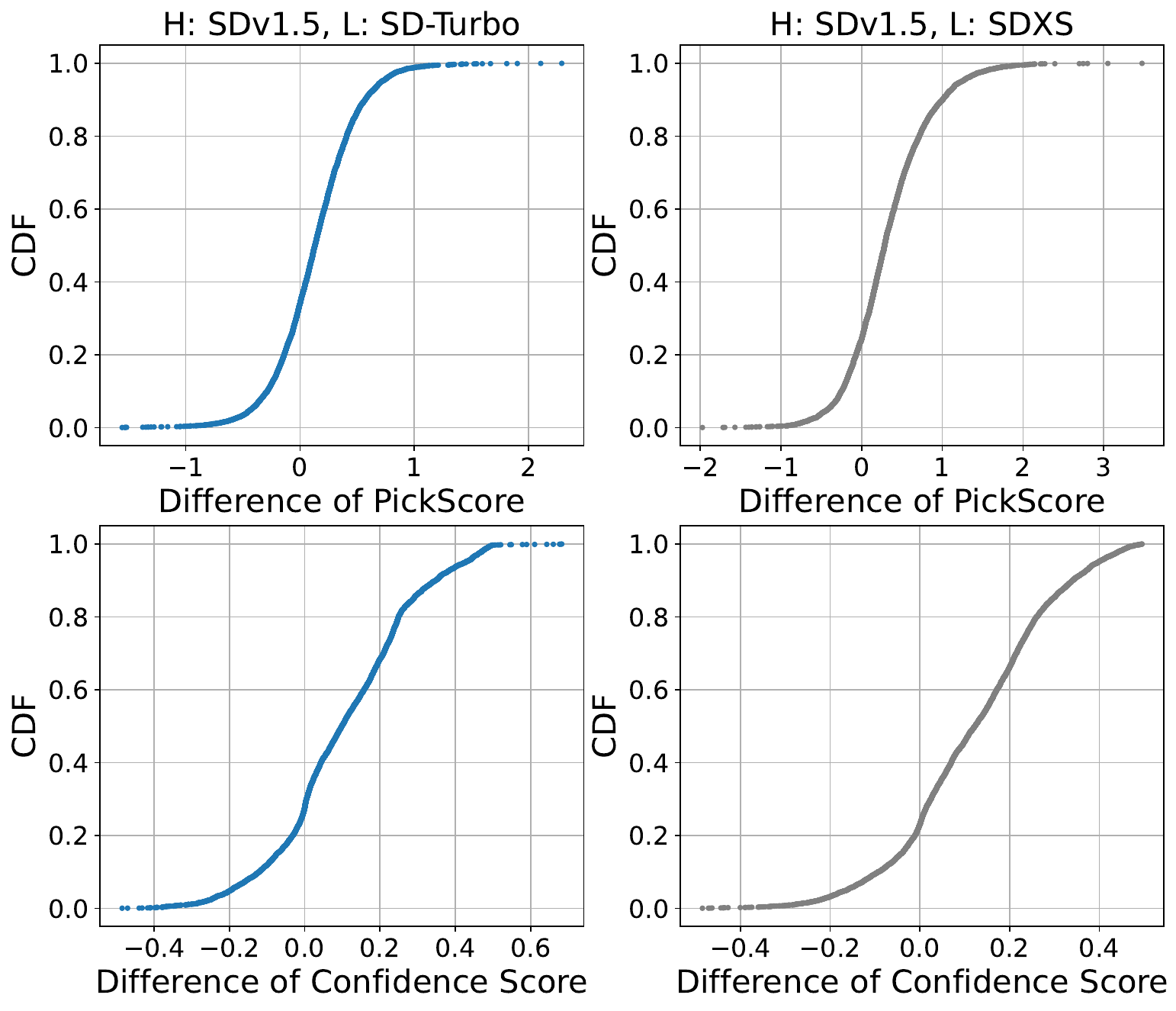}}}%
    % \qquad
    \subfloat[\centering FID vs. serving throughput \label{fig:pareto}]{{\includegraphics[width=.25\textwidth]{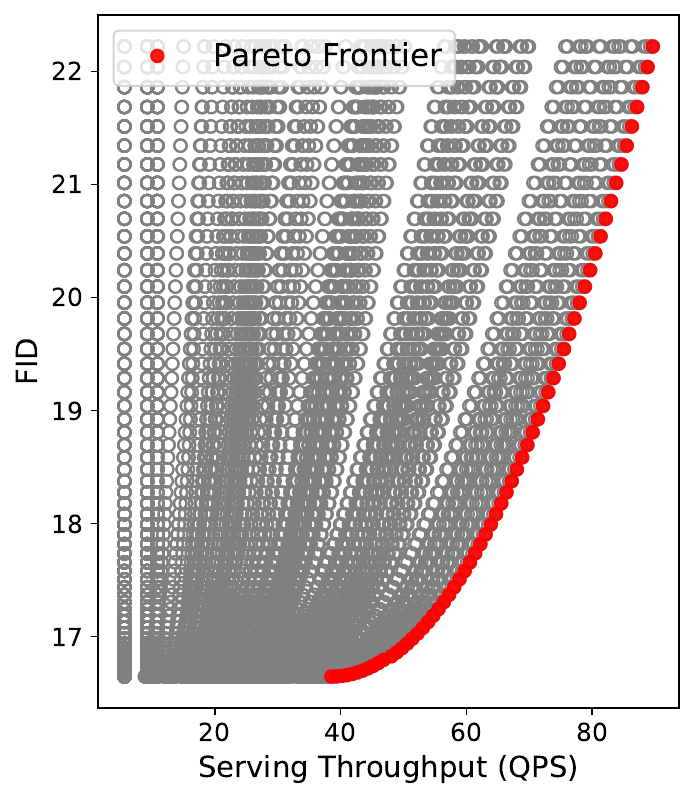}}}%
    \caption{ (a) The quality-latency trade-offs of systems serving independent diffusion models and diffusion model cascades with different discriminator designs with batch size one. The top panel uses diffusion model cascades built with SDv1.5 as the heavyweight model (H) and SD-Turbo as the lightweight model (L). The bottom panel uses SDXS as the lightweight model instead. Lower latency is achieved by using lighter models or treating more queries as easy in diffusion model cascades. 
    FID is the lower the better.  
    (b) The distribution of the difference in image quality between a lightweight model and a heavyweight model. Negative values in the x-axis mean the lightweight model's generated image quality is better than those from the heavyweight model. Top panels use PickScore as the quality metric while bottom panels use confidence score from our proposed discriminator. 
    (c) Illustration of how different resource allocation configurations affect serving throughput (QPS) and response quality (FID). All results use the dataset MS-COCO 2017~\cite{lin2014microsoft}.}
    % \vspace{-1em}
\end{figure*}

% \vspace{-1em}
\subsection{Background and Motivations}
\label{subsec:background_and_motivations}

% \begin{figure}[t]
%   \centering
%     \includegraphics[width=0.9\linewidth]{figures/cascade.pdf}
%   \caption{The quality-latency trade-offs of diffusion models and diffusion model cascades with different discriminator designs. \TODO{replace the figure}}
%   \label{fig:model-variants}
% \end{figure}

Model developers often train diffusion models with varying architectures and sizes to explore trade-offs between model quality and efficiency.
Several quantitative metrics can be used to measure the generated image quality of text-to-image diffusion models, each with its own limitations. 
We highlight a few.  
(1) \textit{Fréchet Inception Distance (FID) score}~\cite{10.5555/3295222.3295408}: it quantifies diffusion model quality by comparing the distribution of generated images with that of real images given a set of prompts. A lower FID score indicates a better model quality. As FID compares two data distributions, it is not suited for assessing the quality of individual images generated from diffusion models.
In this work, we use FID score to quantify the response quality of our serving system given a set of text prompts. 
(2) \textit{PickScore}~\cite{kirstain2023pickapicopendatasetuser}: It compares the quality of images generated using the same text prompt.  
It evaluates alignment relative to each specific prompt, not across varying prompts, making scores incomparable between different prompt-image pairs. We later use PickScore to motivate the existence of easy queries. 
(3) \textit{CLIP Score}~\cite{hessel-etal-2021-clipscore}: It measures the alignment between a text prompt and a generated image from it. 
A higher CLIP score indicates better semantic alignment. 
However, CLIP scores of different model variants can be very close and it does not consistently reflect the image's perceptual quality such as visual realism or aesthetic appeal.

The orange points in Figure~\ref{fig:model-variants} illustrate response quality-latency trade-offs for a system serving diffusion models from HuggingFace~\cite{huggingface_models}, each point representing a model variant. 
These model variants come from off-the-shelf diffusion models with different architectures or the same model executed with a different number of diffusion steps. 
Serving heavier diffusion models can deliver higher-quality images but comes at the cost of higher inference latency and thus lower serving throughput.

Query-aware model scaling is motivated by the observation that certain ``easy'' queries can be processed by lighter diffusion model variants without compromising the quality of the generated images.  
Figure~\ref{fig:easy-hard} illustrates the distribution of the difference in image quality for two light-heavy diffusion model pairs: SD-Turbo vs. SDv1.5 and SDXS vs. SDv1.5. 
As we need a quality metric that compares the image quality from the same text prompt, we use PickScore (top panels) and confidence scores from our discriminator (bottom panels). 
The discriminator design is elaborated in \S\ref{sec:discriminator}. 
The figure shows that, for 20-40\% of the queries (i.e., easy queries), the lightweight model generates images with similar or even better quality than the heavyweight model. 
This observation drives the design of a model serving system that dynamically serves both lightweight and heavyweight model variants, routing easy queries exclusively to the lightweight model to improve system efficiency. 
\subsection{Challenges}
\textbf{Challenge 1: Model Cascading for Diffusion Models.}
One effective approach in identifying easy queries involves constructing diffusion model cascades that combine models of varying sizes, guided by \textit{discriminators}~\cite{viola2001rapid, bolukbasi2017adaptive}. 
The discriminator outputs a \textit{confidence score} indicating whether the generated image meets a predefined quality standard. 
% The optimal discriminator designs are well-established for classification tasks, often relying on the prediction confidence of the small model, typically measured by its softmax probability output~\cite{lebovitz2023efficient, kang2017noscope, chen2020frugalml}. 
 % Conventional metrics used to evaluate diffusion models, such as Fréchet Inception Distance (FID), are not applicable for assessing individual image quality in real time. Other metrics that do assess individual image quality fails to discriminate between easy and hard queries. 

However, designing the discriminator for diffusion models is non-trivial. 
The curves in Figure~\ref{fig:model-variants} show the quality-efficiency trade-offs across various model cascades, each using a unique discriminator design. 
We compare the following approaches: 
(1) \textit{PickScore} and \textit{ClipScore}:  This design leverages widely used quantitative metrics, CLIP Score and PickScore respectively, to assess image quality. If the generated image’s score exceeds a confidence threshold, the query is classified as easy. 
(2) \textit{Random}: This approach uses a random classifier to assign queries, with each query classified as easy with a given probability. 
(3) \textit{Discriminator}: This approach uses our discriminator design detailed in Section~\ref{sec:discriminator}.
The curves are generated by adjusting the confidence threshold for the PickScore, ClipScore, and Discriminator and the probability for the Random. 
For the Random classifier, we conducted the experiments 20 times, with the shaded area 
% in Figure~\ref{fig:model-variants} 
representing standard deviations across these runs.  
% This is because these discriminators fail to reliably identify easy queries, while also adding extra overhead. \TODO{if we think existing metrics cannot identify easy queries, then Figure(b) cannot use pickscore to demonstrate there are easy queries either. Can we only mention the extra overhead? do we have numbers on the overhead compared to the inference latency of the lighweight model ?}
% \TODO{qizheng: Do we still have the line of "discriminator" which is our approach in figure 1a? The paragraph is not talking about our method}

Surprisingly, the results show that discriminators based on established metrics, CLIP Score and PickScore, underperform relative to the baseline Random classifier. This counterintuitive finding shows limitations of these metrics in cascading diffusion models, as they fail to reliably differentiate between easy and difficult queries, resulting in suboptimal routing. In contrast, our discriminator design overcomes these challenges
and outperforms the Random classifier. 

Another surprising observation is that FID gets worse as the latency increases in the end, implying that the overall system response quality can decrease as more queries are routed to the heavyweight model. This is consistent across different light-heavy diffusion pairs. We hypothesize the reason is that including a portion of outputs from the lightweight model yields a more balanced and diverse image representation, which better aligns with the distribution of real images and results in lower FID scores.

\textbf{Challenge 2: Resource Allocation.}
The second challenge lies in effectively leveraging the time savings in processing easy queries to optimize system performance. A model serving system must co-optimize the confidence threshold with other system parameters to maximize serving throughput and response quality while meeting latency deadlines. 
Specifically, a higher confidence threshold imposes a stricter constraint on image quality, leading to better system response quality. 
However, this increased stringency results in a greater proportion of queries being redirected to the heavyweight model, thereby requiring more workers to host it. This increases system loads for a specific query demand and thus decreases the serving throughput of the system. 
On top of these, the batch size to execute the lightweight and heavyweight model affects both their throughput and the overall end-to-end latency of queries, which need to be adjusted accordingly depending on the system loads.   

% \TODO{Sohaib: please revise the text. I want to explain why batch size also need to be adjusted here?}

% In resource-constrained environments, increased system load can result in prolonged query processing times, ultimately resulting in SLO violations.
% Consequently, careful tuning of the confidence threshold is crucial to achieve an optimal balance between accuracy and latency. Effective tuning enables the system to operate within acceptable bounds, mitigating the risk of SLO violations while maintaining satisfactory performance.

% 
Figure~\ref{fig:pareto} illustrates how different resource allocation configurations affect serving throughput and response quality. We cascade SD-Turbo and SDv1.5 using our proposed discriminator and serve it on 10 A100 GPUs. We vary three system configurations: the confidence threshold in the model cascade, batch size, and the model placement.   
Out of all $\sim$9K possible configurations, we are only interested in those at the Pareto frontier, due to the fact that for a given query demand, configurations at the Pareto frontier yield the highest possible response quality compared to other configurations. 
% \TODO{qizheng: fill in the numbers}
% The results imply that identifying the right amount of model scaling requires co-optimizing the confidence threshold, model placement, and batch size.

% \begin{figure}[t]
%   \centering
%     \includegraphics[width=0.59\linewidth]{figures/pareto.png}
%   \caption{Illustration on how different resource allocation configurations affect system throughput (QPS) and response quality (FID). \TODO{change the figure} }
%   \label{fig:pareto}
% \end{figure}

% \TODO{Sohaib: emphasize that the confidence threshold and the rest of the configuration parameters (batch size and model placement) needs to be co-optimized. }

% % \TODO{Queuing model using Little's law}

% Dynamic provisioning: challenges of dynamically provisioning resources to handle varying workloads and different query complexities (i.e., change the confidence threshold)
% Predictive scaling: resource allocation after changing the threshold.
% Resource utilization and efficiency: trade-offs between ensuring SLO requirements and maintaining high utilization of resources.

% \vspace{-1em}
\section{Design of \pjn{}}
\label{sec:design}
We now present the architecture of \pjn{}, a diffusion model serving system that leverages query-aware model scaling to efficiently serve test-to-image diffusion models. 

\begin{figure}[t]
  \centering
  \includegraphics[width=0.99\linewidth]{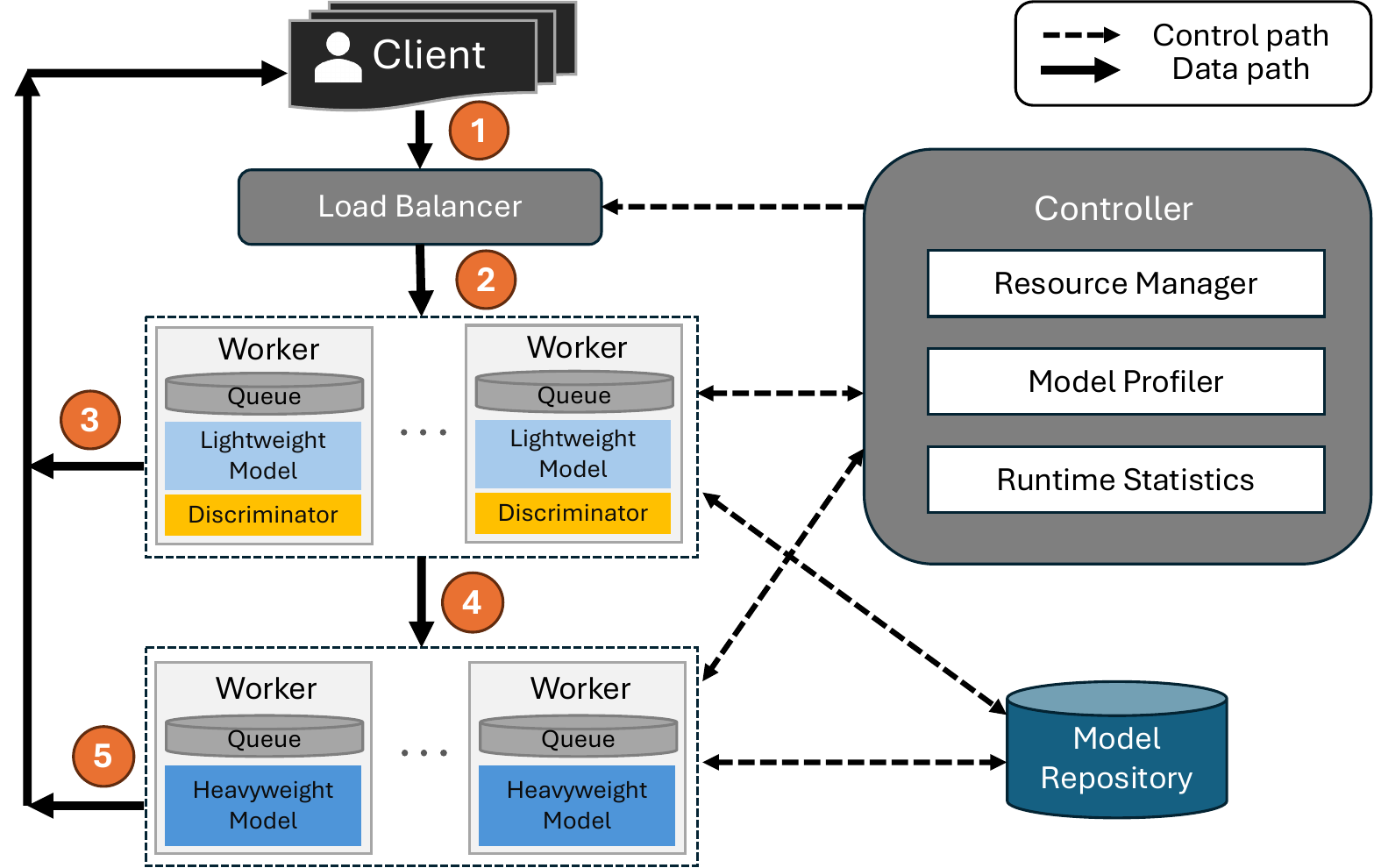}
  % \vspace{-1em}
  \caption{System architecture of \pjn{}: (1) The query from the client is sent to the load balancer, (2) The load balancer sends it to a worker with the lightweight model and the discriminator, (3) If the confidence score is greater than the threshold, the response is sent back to client, (4) Else, the query is sent to a worker with the heavyweight model, and (5) its output is sent to the client.}
  \label{fig:system_architecture}
  % \vspace*{-7mm}
\end{figure}

% \vspace{-1em}
\subsection{Overview}
\label{subsec:design_overview}
Figure~\ref{fig:system_architecture} shows the system architecture. It has separate data and control paths. In the data path, clients send queries to the system's Load Balancer, which routes them to suitable workers and returns generated images. 
In the control path, the Controller periodically re-allocates resources depending on the system runtime statistics collected from the workers. 
% In the following, we explain the main components of the system.
% and how the system adapt to fluctuating query demands.  

\textbf{Controller.} The Controller manages the resources in the system. It uses the Resource Manager to allocate a model variant to each worker and set its batch size and the confidence threshold for the workers hosting the lightweight model and its discriminator. 
It periodically collects runtime information from the workers to update model execution profiles as well as the queue lengths and demands seen by each of the workers to inform resource allocation decisions.

\textbf{Model Repository.} It manages the registration of diffusion model variants and hosts these registered variants, along with the discriminators used to cascade between them. 

\textbf{Load Balancer.} The Load Balancer sits on the data path between a client and workers. Upon receiving queries from clients, the Load Balancer initially routes each query to a worker running a lightweight diffusion model. 
If the generated image’s quality estimated by the discriminator meets the quality requirement, specified as a confidence threshold, it is returned to the Load Balancer as the response. 
Otherwise, the query is forwarded to a worker hosting the heavyweight diffusion model to generate the final response. 
% \TODO{Sohaib: Please revise the text. Also, Please see if you would like to clarify these in the text too: does the load balancer route queries in round robin fashion? when routing a query to the heavyweight model, does lood balancer do it or the worker which executes the lightweight model?} 

\textbf{Workers.} Each worker executes its hosted model variant to serve queries routed to it and kept in its local queue. Some workers host the lightweight models together with the discriminators while the rest host the heavyweight models. The batch size, which model variant to host, and the confidence threshold for each worker are determined by the Controller. 

% \TODO{Sohaib: I remove your query flow etc content as the query flow should be clear after discussing the load balancer and the worker. We can add a paragraph here explaining how the system adapt to fluctuating query demand here. But it is also kinda duplicates the paragraphs in Section 2 Challenge 2.}

We next explain the two technical innovations of \pjn{}, the discriminator design for cascading diffusion models (\S\ref{sec:discriminator}) and the resource allocation algorithm (\S\ref{sec:allocation}).

\subsection{Discriminator Design for Model Cascading} 
\label{sec:discriminator}
\begin{figure}[t]
  \centering
  \includegraphics[width=0.99\linewidth]{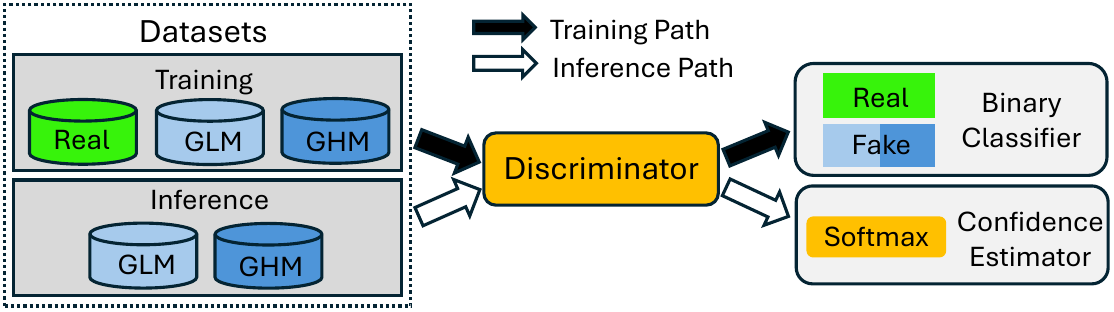}
  % \vspace{-1em}
  \caption{Training and inference paths of the discriminator. `Real' refers to images sourced from real-world high-quality datasets. `Fake' refers to the generated images from diffusion models. GLM: \underline{G}enerated images from \underline{L}ightweight diffusion \underline{M}odel; GHM: \underline{G}enerated images from \underline{H}eavyweight diffusion \underline{M}odel. }
  \label{fig:discriminator_path}
  % \vspace{-1.5em}
\end{figure}

% Model cascading is a structured approach that leverages multiple models of varying complexity to efficiently process incoming queries. In this system, all models are diffusion models, but they differ in the number of diffusion steps. 
% A lightweight model with fewer diffusion steps generates the initial image quickly, optimizing for response time. However, if the image does not meet the required quality standards, the request is redirected to a heavyweight model with more diffusion steps, which generates a higher-quality image but requires more time. 

At the core of a diffusion model cascade is the discriminator, which evaluates the quality of images generated by a diffusion model to determine whether deferral to a heavyweight model is necessary. The discriminator must be accurate in quality estimation and efficient to minimize runtime overhead. 
Our approach leverages the insight that an ML model can be trained to accurately distinguish between images generated by diffusion models and real images. 
This ML model can then be repurposed as the discriminator to differentiate whether the images produced by the lightweight model meet quality requirements based on its classification confidence.

\textit{Discriminator Design.}
Figure~\ref{fig:discriminator_path} illustrates the offline training process for preparing the discriminator and the inference process for its use within the diffusion model cascade.  The discriminator is trained on a binary classification task to distinguish between high-quality, real-world images (labeled as `real') and generated images (labeled as `fake').
Real images are sourced from datasets like MSCOCO~\cite{lin2014microsoft} and DiffusionDB~\cite{wangDiffusionDBLargescalePrompt2022}, while generated images come from both the lightweight and heavyweight diffusion models.
% Through this training, the discriminator learns to classify images as either "real" (from the dataset) or "fake" (generated by models). 
This task enables the discriminator to detect visual differences between high-quality real images and generated ones, including variations in sharpness, texture coherence, and artifact presence, equipping it to assess image quality accurately within the cascade framework.

% The training dataset for the discriminator is a combination of \textit{real images} from datasets like MSCOCO~\cite{lin2014microsoft} and DiffusionDB~\cite{wangDiffusionDBLargescalePrompt2022}, and \textit{fake images} generated by both the lightweight and heavyweight diffusion models. 
% The discriminator learns to classify whether a given image is "real" (from the dataset) or "fake" (generated by the models). This binary classification task equips the discriminator with the ability to detect visual discrepancies between real-world images and generated images, such as differences in sharpness, texture coherence, and the presence of artifacts.

During inference, the discriminator receives the image produced by the lightweight model and outputs a softmax value between 0 and 1, representing the likelihood that the image belongs to the `real' or `fake' class. This value is referred to as the \textit{confidence score}, representing how likely the input image is to resemble a `real' image, implying higher quality. 
Our implementation uses EfficientNet as the discriminator architecture because it has much lower computation complexity compared to even the lightweight diffusion model variants while achieving high classification accuracy. Section~\ref{sec:eval_cascading} evaluates the impact of alternative design choices.

\textit{Model Cascading.} Using the discriminator, cascading a light-heavy diffusion model pair is straightforward by setting a \textit{confidence threshold} to quantify image quality requirements.
If the confidence score of an image for a query exceeds the specified confidence threshold, the image is returned to the user, as it meets the quality standard. 

% Careful tuning of the confidence threshold is crucial to achieve an optimal balance between image quality and latency of the model cascade. \pjn{} optimizes the confidence threshold with the rest of system parameters to optimize model serving efficiency as described next.  

% When the lightweight model's output receives a confidence score below the threshold, the system forwards the query to the heavyweight model, which employs more diffusion steps to generate a new, higher-quality image. This cascading process ensures that users receive high-quality images efficiently, with the lightweight model handling the majority of queries quickly and the heavyweight model only being involved when necessay.

% \hui{When there are more than two model variants, do we consider model cascading with all of these models? How do we decide the best depth of the model cascading and the model variant to use for at each stage of the cascaded pipeline?}

% Introduce the 2-level cascaded serving system. Currently we only use few-step diffusion models in the first level and the best normal diffusion models with the highest scores (FID) in the second level, such that the latency of the first level can be relatively small compared to the second level. Quantitatively measuring the quality of images can be challenging, especially when there are more than two levels. When there are more levels, the additional overhead can be also non-negligible for diffusion model serving system. 

% \vspace{-1em}
\subsection{Resource Manager}
\label{sec:allocation}

The Resource Manager dynamically adjusts the confidence threshold, allocates models across servers in a cluster, and configures batch sizes to respond to varying query demands. 
By tuning the confidence threshold, the model serving system leverages the quality-latency trade-off inherent in model cascades to adapt efficiently. For instance, during periods of low demand, a higher threshold prioritizes image quality, while at peak times, a lower threshold ensures latency deadlines are met by allowing minor quality compromises.

This model scaling approach enables a cluster to manage high query volumes gracefully, avoiding overload while tolerating minor reductions in response quality as necessary. 
Unlike traditional resource provisioning, which typically dedicates resources based on peak demands and results in underutilization during off-peak times, model scaling optimizes resource usage across fluctuating loads. Previous work Proteus~\cite{ahmad2024proteus} explored model scaling by selecting appropriate model variants to host based on loads. However, model cascades introduce dependencies between model variants for query-aware processing based on query complexities, necessitating a new resource allocation algorithm to manage these dependencies effectively.   

Our resource allocation algorithm centers on building performance models for three key performance metrics: {\em serving throughput}, {\em latency}, and {\em response quality}. With these models, identifying the optimal resource allocation can be stated as an optimization problem and formulated within a mixed integer linear programming (MILP) framework, allowing for efficient solutions using MILP solvers. Response quality is directly influenced by the confidence threshold, denoted as 
$t$. Below, we describe the constraints on latency and serving throughput and then our resource allocation formulation.
 
\textit{Latency Constraints.} Model dependencies in model cascades introduce complexities in estimating the total time a query spends in the system. 
To keep query latency within the defined SLO, the Resource Allocation component must consider two main factors: (i) execution latency and (ii) queuing delays. 
Appropriate batch size settings for each model are essential for managing these latencies. While larger batch sizes increase throughput, they also raise the execution latency for each query within the batch. 
As the execution time of text-to-prompt diffusion models is highly deterministic, execution latency can be accurately predicted and profiled across different batch sizes.

To estimate queuing delays for each model, we apply Little's law~\cite{shortle2018fundamentals}, i.e. $W = \frac{L}{\lambda}$, where $W$ is the waiting or queuing time, $L$ is the length of the queue, and $\lambda$ is the query arrival rate. As mentioned in Section~\ref{subsec:design_overview}, the Controller maintains a record of the queue length for each worker and the demand seen by each worker. Using these values, we can estimate the queuing delay for each model. 

% \TODO{Sohaib: Please revise the paragraph to clarify the following question: based on the description, we will have a queueing model separately for each worker. How do we decide a single batch size for the lighweight model and for a heavyeweight model that are hosted on different worklers?}

Mathematically, let $b_1$ and $b_2$ be the batch size used to execute the light and heavy models respectively. let $e(.)$ measures the execution latency and $q(.)$ models the queuing delays. The latency a query experienced in the system should be less than the deadline in SLO requirement $L$: 
% \vspace{-0.5em}
\begin{equation}
    e(b_1) + q(b_1) + e(b_2) + q(b_2) \leq L \label{eq:latency}
\end{equation}

\textit{Throughput Constraints.} The system's serving throughput can be limited by the throughput of workers hosting either the lightweight or heavyweight model. Mathematically, let $x_i$ represent the number of devices allocated to serve the $i$-th model variant and $T_i(.)$  represents the throughput of a single worker for the $i$-th model variant, collected through profiling. Here, $i=1$ refers to the light model and $i=2$ refers to the heavy model. The serving throughput of each model variant in the model cascade must meet or exceed their respective estimated query demands:  
% \vspace{-0.5em}
\begin{align}
    x_1 . T_1(b_1) &\geq D, \\
    x_2 . T_2(b_2) &\geq D . f(t), \\
    x_1 + x_2 &\leq S, \label{eq:throughput}
\end{align}
where $D$ is the total estimated query demand entering the system (which is also the demand for the lightweight model), $f(t)$ represents the fraction of queries deferred to the heavyweight model when the confidence threshold is set to $t$, and $S$ denote the total number of devices available to the system. $f(t)$ is initialized through offline profiling and updated during model serving as $t$ changes. 

\textit{The resource allocation problem.} We formulate an optimization problem that maximizes the confidence threshold given incoming query demand. 
The optimization is a mixed integer linear program (MILP) that tunes server count allocated for each model variant ($x_1$, $x_2$) and batch sizes ($b_1$, $b_2$).
% \vspace{-1em}
\begin{align}
    \max_{x_1,x_2,b_1,b_2} & t \\
    s.t. \quad & \text{Constraints Eq. \ref{eq:latency} - \ref{eq:throughput}} \nonumber 
\end{align}
% \vspace{-1em}

\textit{Solving the MILP.} 
The optimization problem is solved periodically by invoking a MILP solver. We estimate query demand $D$ using an exponentially weighted moving average on demand history. To accommodate micro-scale variations in query arrivals, we use $\lambda D$ as the estimated query demand in Eq.~\ref{eq:throughput}, where $\lambda$ is the over-provisioning factor and set to 1.05 by default. The time overhead to solve the MILP does not lie on the critical path of query processing as the MILP is called asynchronously and its execution is in the control path. We compare the resource allocation algorithm with alternatives and report its runtime overhead in Section~\ref{sec:eval_allocation}.

% \TODO{What is $f(t)$? How do we measure or model it?}

% \TODO{How do we get throughput? Isn't it just a function of execution latency?}

% \TODO{How do we get demand? Explain that it is measured using an exponentially weighted moving average.}

% Details of confidence threshold changing with workloads, dynamic resource allocation for each stages, adaptive batching scheduling.

% \subsection{Confidence-based Reuse Optimization}
% Do reuse, or refinement, only to improve the image quality, instead of reducing the latency. For all queries that go to the second stage, generate new images based on both the texts and images from the first stage, since reusing initial images typically gives better images compared with generating images from scratch in terms of FID, Clip Score, and PickScore. 

% A brief explanation on why we do not reduce number of steps for lower latency by reusing images: current metrics are not so sensitive to the change of number of denoising steps when we apply reusing (the plots showing FID/PickScore/Clip changes with different number of steps when doing reuse), thus we can naively use fewer steps (e.g., 40, 30, etc). But it is not easy to predict how many steps are needed given a query and corresponding images.

% \vspace{-1em}
\section{Evaluation}
This section evaluates the efficacy of \pjn{} by answering the following questions: 
\textbf{Q1}: How does the performance of \pjn{} compare to alternative approaches on both synthetic traces (\S\ref{sec:synthetic}) and real-world traces (\S\ref{sec:real})?
\textbf{Q2}: How does the model cascading design (\S\ref{sec:eval_cascading}) and the resource allocation algorithm (\S\ref{sec:eval_allocation}) compare to alternative approaches? What is the runtime overhead of the MILP solvers? 
\textbf{Q3}: How do different SLO settings affect the performance of \pjn{} (\S\ref{sec:sensitivity})?

% \vspace{-1em}
\subsection{Experiment Settings}

\textbf{Implementation of \pjn{}.} We implement \pjn{} in a simulator and on a testbed cluster. (1) \textit{The simulator-based implementation} consists $\sim$7K lines of Python code. It uses an event queue and a timer to record the arrival and processing of queries. The execution time of queries for the diffusion models is profiled for offline usage. 
(2) \textit{The cluster-based implementation} aims to test the performance of \pjn{} on actual GPU hardware. We use the HuggingFace~\cite{von-platen-etal-2022-diffusers} and PyTorch frameworks~\cite{paszke2019pytorchimperativestylehighperformance} to execute the diffusion models and discriminator for inference. 
Our cluster consists of 16 NVIDIA A100 GPU workers, and we use gPRC for fast and lightweight communication between system components such as the Controller and Workers.
The results we reported in the paper are collected from the simulator unless noted differently. 
We later show that the results from our simulator closely match the results from our cluster-based implementation, with a slight discrepancy caused by variance in processing queries on actual GPUs. The detailed difference is reported in Section~\ref{sec:real}. We use Gurobi~\cite{gurobi} to solve our MILP optimization.

% \TODO{Difference in cluster and simulation results (quantitative)}
% \TODO{Which experiments use simulator and which use testbed?}

% \TODO{Model switching explanation?}

\textbf{Diffusion Models and Datasets.}
We construct model cascades using three light-heavy diffusion model pairs. These different diffusion pairs aim to show the generalizability of the proposed discriminator design and the effectiveness of \pjn{} across different model configurations. 
\textbf{Cascade 1:} We use SD-Turbo~\cite{sauer2023adversarialdiffusiondistillation} as the lightweight model and SDv1.5~\cite{Rombach_2022_CVPR} as the heavymodel. SD-Turbo is a fast generative text-to-image model that can generate an image from a prompt in only one step. We use SDv1.5 with 50 steps. 
The inference latency to generate one image for a text prompt on A100-80GB for SDv1.5 and SD-Turbo is $\sim$1.78s and $\sim$0.1s respectively. We set the SLO for this Cascade to be 5s for our experiments and explore the effect of different SLO values on performance in \S\ref{sec:sensitivity}. 
\textbf{Cascade 2:} We use SDXS-512-0.9 (referred to as SDXS)~\cite{song2024sdxsrealtimeonesteplatent} as the lightweight model and SDv1.5 as the heavymodel. SDXS takes $\sim$0.05s to generate an image from a prompt in one step on an A100-80G GPU. We again use an SLO of 5s for this Cascade.
\textbf{Cascade 3:} We use SDXL-Lightning~\cite{lin2024sdxllightningprogressiveadversarialdiffusion} with two steps ($\sim$0.5s to generate an image for a prompt) as the lightweight model and SDXL~\cite{podell2023sdxl} with 50 steps ($\sim$6s for the same generation) as the heavyweight model. As this cascade is heavier, we use an SLO of 15s.

For datasets, we use \textit{MS-COCO 2017}~\cite{lin2014microsoft} for Cascades {1}-{2} which generate images at a resolution of 512x512, and \textit{DiffusionDB}~\cite{wangDiffusionDBLargescalePrompt2022} for Cascade~{3} which generates images at a resolution of 1024x1024. 
We select the first 5K text-image pairs from each dataset with text prompts serving as queries and images applied in calculating FID scores for evaluation. Further, we use the Microsoft Azure Functions trace~\cite{shahrad2020serverless} as a representative real-world workload to drive load on the system. We scale the trace using shape-preserving transformations to match the capacity of our system.

\textbf{Evaluation Metrics.} We access system performance using two key metrics. (1) Response quality (\textbf{FID}): FID measures the similarity between two distributions -- generated images and ground truth images. To compute the FID score for a given system configuration, we process all text prompts in a dataset through the system and evaluate the quality of the generated images. 
(2) \textbf{SLO Violation Ratio}: This metric represents the proportion of queries that fail to meet the SLO latency requirement or are preemptively dropped by the system when they are predicted to miss the deadline.
We vary the query demand (in QPS) entering the system and report how changes in demand affect system response quality and the SLO violation ratio.

% \TODO{qizheng: We calculate the FID with 20k images generated by diffusion models. Among the images, we calculate the FID with different ratio of images coming from lightweight model only and the other coming from heavyweight model. Then during the serving time, at each second, we compute the ratio (\# queries go to heavyweight / \# all queries received), and map it to get the pre-computed FID. We need to explain how we calculate the FID scores, as we mentioned previously that FID is not used to evaluate an individual image. That also helps to explain why all-at-large and all-at-small are flat lines for FID, as it should change when we get different queries are different times.}

    % \item \textbf{SLO violation ratio}. The SLO (Service Level Objective) violation ratio 
    % \item \textbf{Query Per Second (QPS)} \TODO{Fill this in}

\begin{table}[t]
\centering
\begin{tabular}{|l||c|c|}
\hline
\textbf{Approach} & \textbf{Allocation} & \textbf{Query-aware} \\ \hline
Clipper-Light & \textcolor{red}{Static} & \textcolor{red}{No} \\
Clipper-Heavy & \textcolor{red}{Static} & \textcolor{red}{No} \\
Proteus & \textcolor{teal}{Dynamic} & \textcolor{red}{No} \\
\pjn{}-Static & \textcolor{red}{Static} & \textcolor{teal}{Yes} \\
\textsc{DiffServe} & \textcolor{teal}{Dynamic} & \textcolor{teal}{Yes} \\ \hline
\end{tabular}
\caption{Comparison of \textsc{DiffServe} with baselines}
\label{table:baselines}
% \vspace{-1em}
\end{table}

\textbf{Counterparts for Comparison.}
We evaluate \pjn{} against four approaches. Table~\ref{table:baselines} highlights their difference.  
\begin{itemize}[noitemsep, nolistsep]
    \item \textbf{Clipper-Light} and \textbf{Clipper-Heavy} are static baselines that route all queries to the lightweight diffusion model and the heavyweight diffusion model, respectively. They implement Clipper~\cite{clipper} to serve these models. Although we use Clipper, this baseline is also representative of other static and query-agnostic model serving systems, such as TensorFlow-Serving~\cite{tfserving}.

    \item \textbf{Proteus}~\cite{ahmad2024proteus} is a model serving system that dynamically selects models based on changing query demand. However, its query routing strategy does not account for query complexity; instead, it randomly assigns incoming queries to model variants, disregarding the content or difficulty of each query.

    \item \textbf{\pjn{}-Static} is a variant of our system that uses a model cascade with a discriminator to estimate query difficulty and route queries appropriately. It is static as it is provisioned for peak and does not adapt confidence threshold to changing system demand. We consider this a practical baseline as it reflects a common practice in production systems where resources are provisioned to accommodate maximum anticipated demand.
\end{itemize}

% \vspace{-1em}
\subsection{Performance Comparison on Synthetic Traces}
\label{sec:synthetic}

\begin{figure*}[t]
  \centering
    \includegraphics[width=0.3\linewidth]{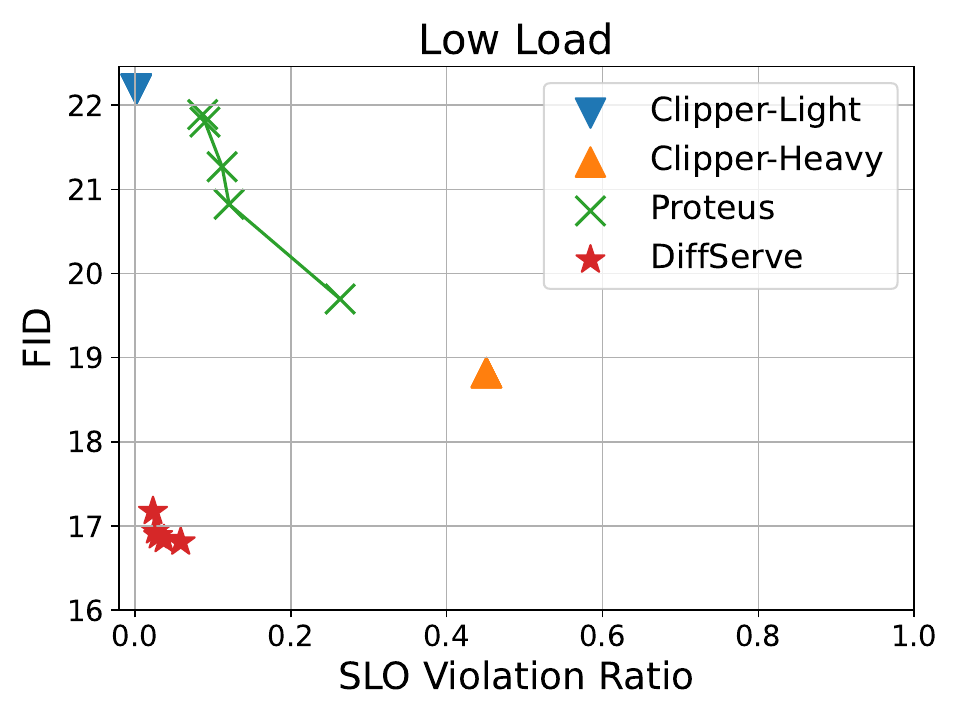}
    \includegraphics[width=0.3\linewidth]{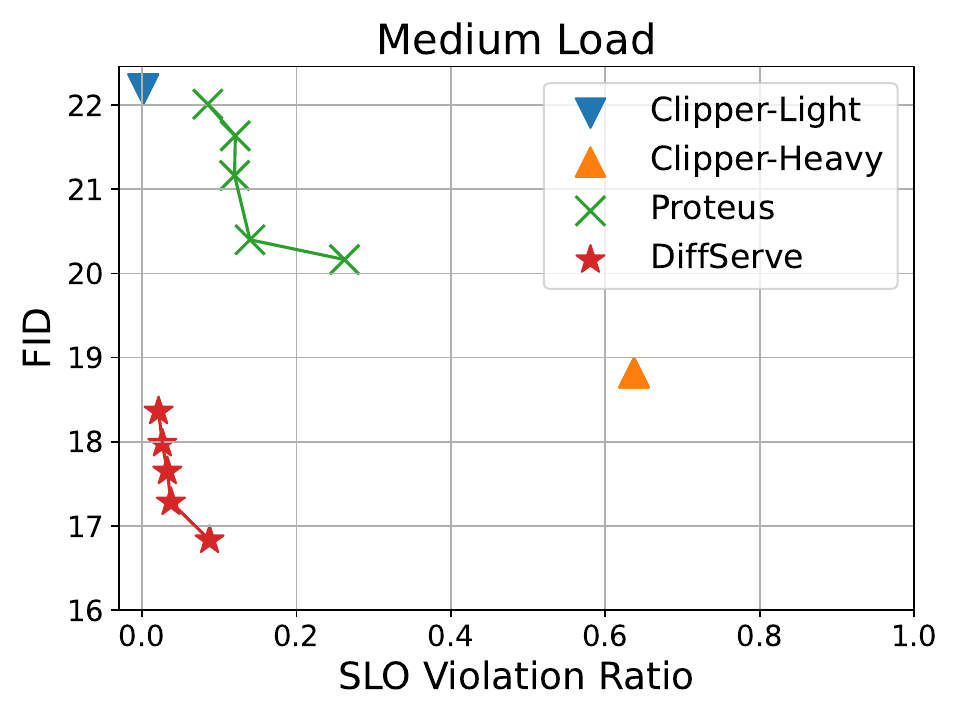}
    \includegraphics[width=0.3\linewidth]{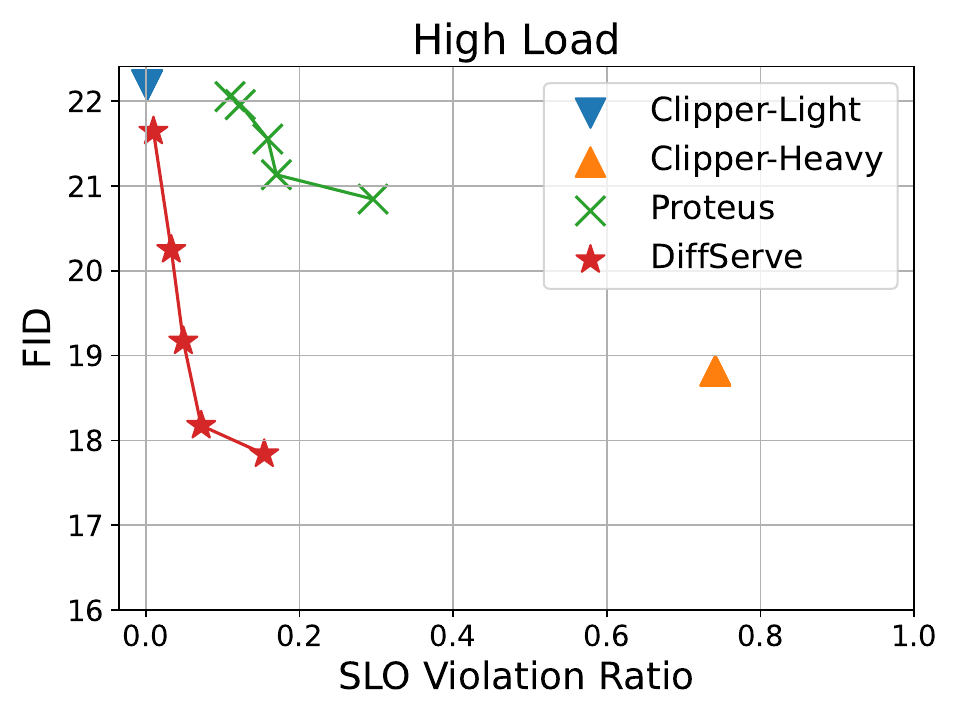}
    % \vspace{-1em}
  \caption{Performance comparison on static trace. \pjn{} offers Pareto optimal between FID and SLO violations (lower left curve).
  }
  \label{fig:baseline_static_comparison}
  % \vspace{-1em}
\end{figure*}

Figure~\ref{fig:baseline_static_comparison} shows the performance of \pjn{} against the baselines on synthetic, static traces for Cascade 1. We vary the load from low to high and observe the effect on all approaches. 
Under static query demand, \pjn{}-Static and \pjn{} perform identically, as there is no need for dynamic adjustment of the confidence threshold. To generate multiple performance points under static query demand, we vary the over-provisioning factor (see \S\ref{sec:allocation}) for both the dynamic approaches, i.e., Proteus and \pjn{}, to explore the quality-latency trade-off.
The static approaches, i.e., Clipper-Light and Clipper-Heavy only offer a single point of performance for each graph since they cannot be tuned to navigate this quality-latency trade-off.

We see that \pjn{} offers the Pareto optimal trade-off between the FID (i.e., response quality) and SLO violations across all three levels of load. 
Although Clipper-Light achieves the smallest SLO violations because it serves the lightweight model for all queries, it suffers from poor response quality. 
The Clipper-Heavy counterpart offers better quality as it uses the heavyweight model for all queries, but has the highest SLO violations (45.10\%-74.11\%) across all approaches. Proteus offers a middle-ground performance between the two Clipper extremes by tuning the model variant based on demand. However, since it routes queries to the model variants randomly instead of considering query content, it still offers sub-optimal performance. As we can see from the Figure, \pjn{} offers the Pareto optimal between FID and SLO violations as the \pjn{} curve lies on the lower left portion of the graph across all loads. We attribute this to the query-aware nature of \pjn{}, as it can select the appropriate model for each query based on its content and difficulty. As we show in \S\ref{sec:eval_cascading}, the overhead of estimating the query difficulty is negligible.

Note that \pjn{} can even outperform Clipper-Heavy in terms of FID because, as shown in \S\ref{subsec:background_and_motivations}, the lightweight model offers similar or better quality for 20-40\% of the queries. This phenomenon is also reported by ~\cite{ding2024hybridllmcostefficientqualityaware} for LLMs, where small LLMs outperform large LLMs for certain queries. This further emphasizes the importance of query-awareness for model selection.
% This difference is the highest for the Low Load condition, and becomes smaller as the load increases \TODO{Why?}

\begin{figure}[t]
  \centering
    \includegraphics[width=0.75\linewidth]{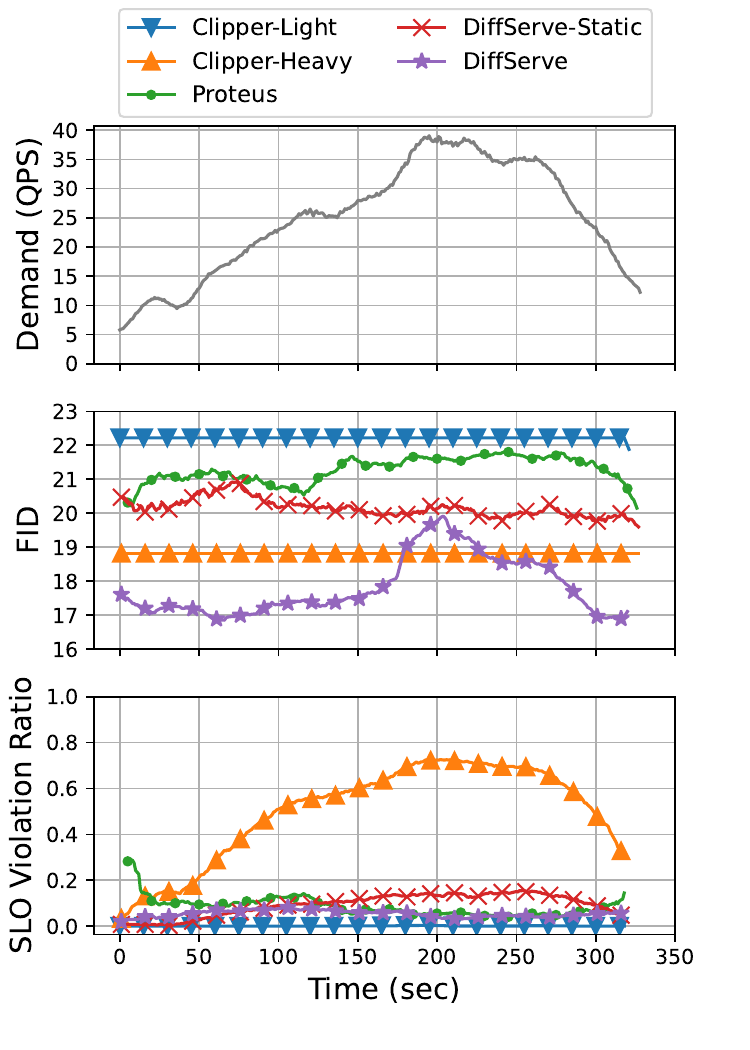}
    % \vspace{-.4cm}
  \caption{Performance comparison on real-world trace for Cascade 1. \pjn{} improves quality by up to 23.4\% over baselines while maintaining low SLO violations. During peak, it offers similar or better quality than static approaches with significantly lower SLO violations (from 19-70\%).}
  \label{fig:baseline_trace_comparison}
  % \vspace{-1em}
\end{figure}

\begin{figure}[t]
  \centering
    \includegraphics[width=0.49\linewidth]{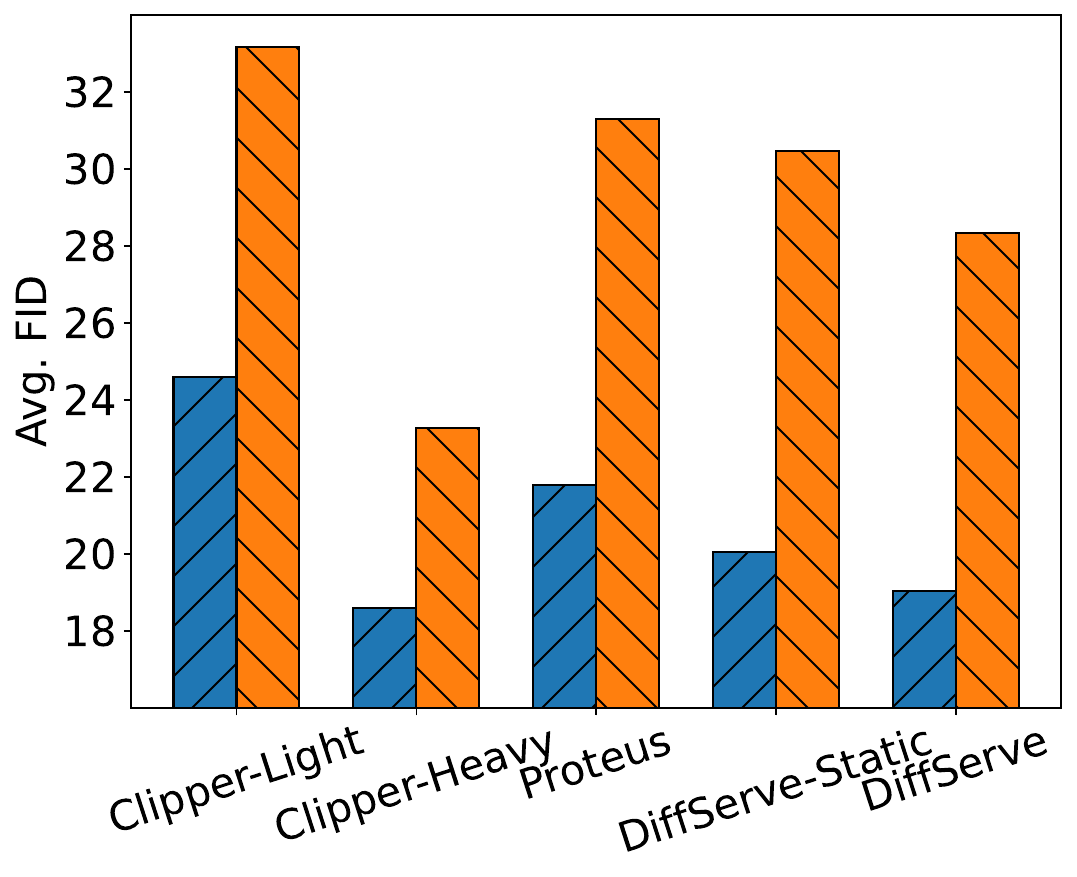}
    \includegraphics[width=0.49\linewidth]{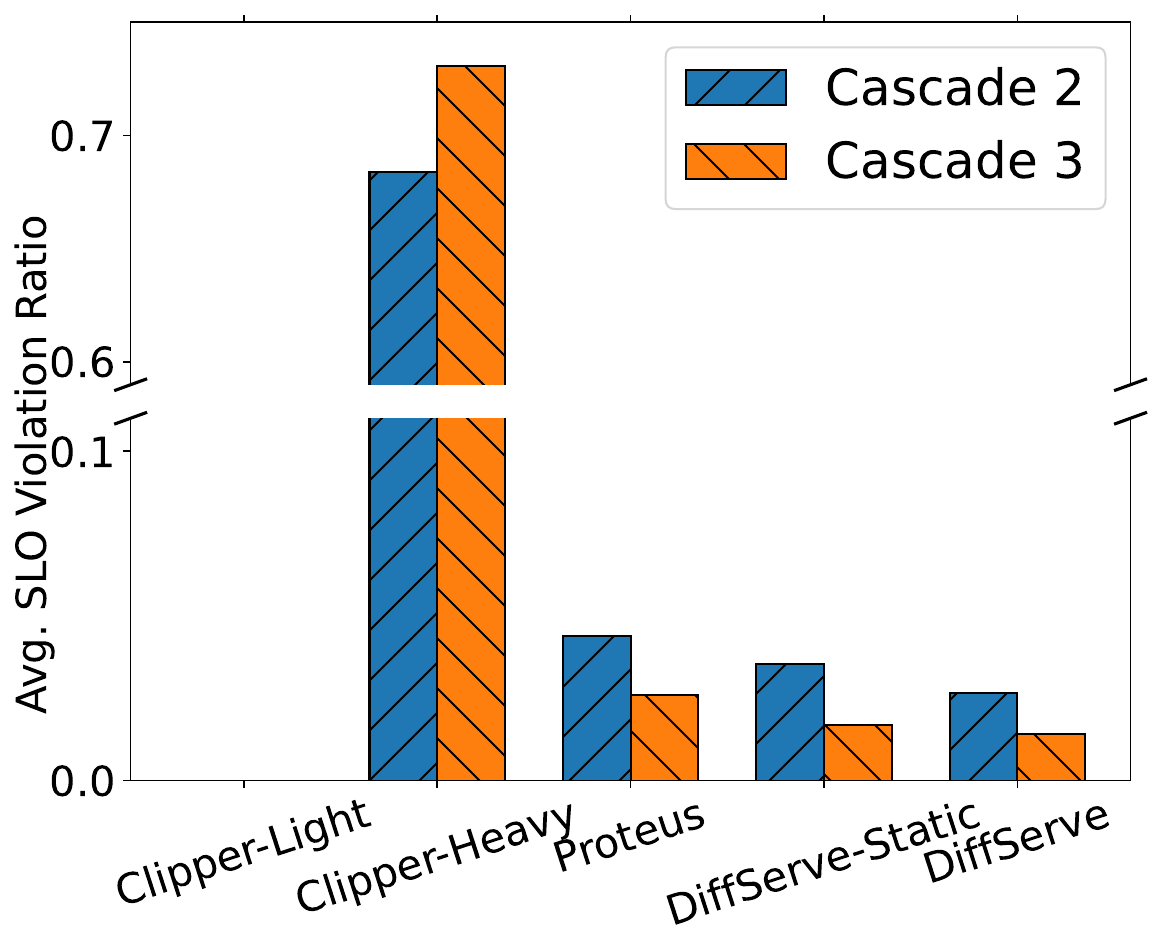}
    % \vspace{-1em}
  \caption{Comparison of approaches on the testbed. Across both Cascade 2 and 3, \pjn{} achieves an 6\%-24\% reduction in average FID compared to all baselines, except Clipper-Heavy which yields significantly high SLO violations. \pjn{} reduces the average SLO violation by up to 1.4$\times$, 1.9$\times$ and 52$\times$ compared to \pjn{}-Static, Proteus and Clipper-Heavy. Clipper-Light's avg. SLO Violation Ratio is zero.}
  \label{fig:baseline_trace_cascade23}
  % \vspace{-1em}
\end{figure}

% \vspace{-1em}
\subsection{Performance Comparison on Real Traces}
\label{sec:real}
% \TODO{added new figure for cascade 2 and 3. Need to mention we scale down the trace for cascade 3 as the execution time for sdxl is long}

Production systems observe significant variations in demand throughout the day. An approach that dynamically responds to demand variations can offer significant performance improvements over static approaches. We now show how \pjn{} performs on a real-world dynamic trace against the baselines. 

Figure~\ref{fig:baseline_trace_comparison} reports the comparison using Cascade~1.
We again observe that Clipper-Light has the lowest quality (highest FID) and low SLO violations due to sending all queries to the lightweight model. Clipper-Heavy offers higher quality ($\sim$15\%) but suffers from significant SLO violations at the peak (up to 75\%) as the heavyweight model has a long execution time. Proteus dynamically tunes the resource allocation according to demand changes and thus experiences an almost consistent level of quality throughout the trace. 
However, as it is query-agnostic, its quality improvement over Clipper-Light is minimal ($<$5\%). We also study the performance of \pjn{}-Static in this case. As it is query-aware but static, it also experiences a consistent level of quality (FID) throughout the trace. Due to its query-aware nature, it can improve quality over Clipper-Light and Proteus by up to 9\%. However, as it cannot change its allocation to accommodate demand changes, it suffers from high SLO violations (up to 19\%) during the peak.

\pjn{} offers the best performance throughout due to its dynamic resource allocation and query-awareness. During the off-peak, it can significantly improve quality (up to 23.4\%) while guaranteeing very low SLO violations. It does so by using the heavyweight model intelligently, only sending it queries that have a low confidence score from the lightweight model, thus ensuring high quality while maintaining low latency. As demand increases and resources get constrained, it keeps SLO violations low by routing more queries to the lightweight model. At peak demand, its FID score is momentarily worse only than Clipper-Heavy, but as mentioned before, Clipper-Heavy suffers from significantly high SLO violations at this time. Therefore, \pjn{} adapts dynamically to the real-world trace and outperforms baselines in terms of both quality and SLO violations.
Under imbalanced workloads dominated by easy or difficult queries, while resources would shift toward lightweight or heavyweight models respectively, \pjn{} would still balance response quality and SLO violations by tuning the confidence threshold.

Figure~\ref{fig:baseline_trace_cascade23} presents testbed results for Cascades 2 and 3. As both configurations exhibit similar trends to Cascade 1, we report the average FID scores and SLO violation rates. We observe that \pjn{} reduces average FID scores by 6\%-24\% for Cascade 2 and by 8\%-15\% for Cascade 3 compared to all other baselines, except Clipper-Heavy, which incurs a high SLO violation (68.4\% and 73\%). 
\pjn{} outperforms all baselines in terms of SLO violations with $1.4\times$, $1.7\times$, and $26\times$ lower violation ratio than \pjn{}-Static, Proteus, and Clipper-Heavy, respectively, for Cascade 2, and $1.2\times$, $1.9\times$, and $52\times$ lower violation ratio for Cascade 3.
% While \pjn{} achieves a lower average SLO violation rate than others, it is 1\% higher than \pjn{}-Static. This is because the static approach provisions for peak demand, keeping off-peak SLO violation very low and lowering overall average.

% End-to-end performance: accuracy changes over time, utilization, SLO violation, etc for each cascade configurations and baselines.

We conducted the same experiments on the simulator, observing an average difference of only 0.56\% for FID and 1.1\% for SLO violations compared to the testbed. This close alignment between the simulator and testbed results confirms the simulator's reliability. Therefore, we use the simulator for the remaining subsections to efficiently evaluate \pjn{} across a broad range of scenarios.

\subsection{Evaluation of the Discriminator Design}
\label{sec:eval_cascading}

\begin{figure}[t]
    \centering
    \subfloat[\centering SD-Turbo \label{fig:discriminator-sdturbo}]{{\includegraphics[width=.51\linewidth]{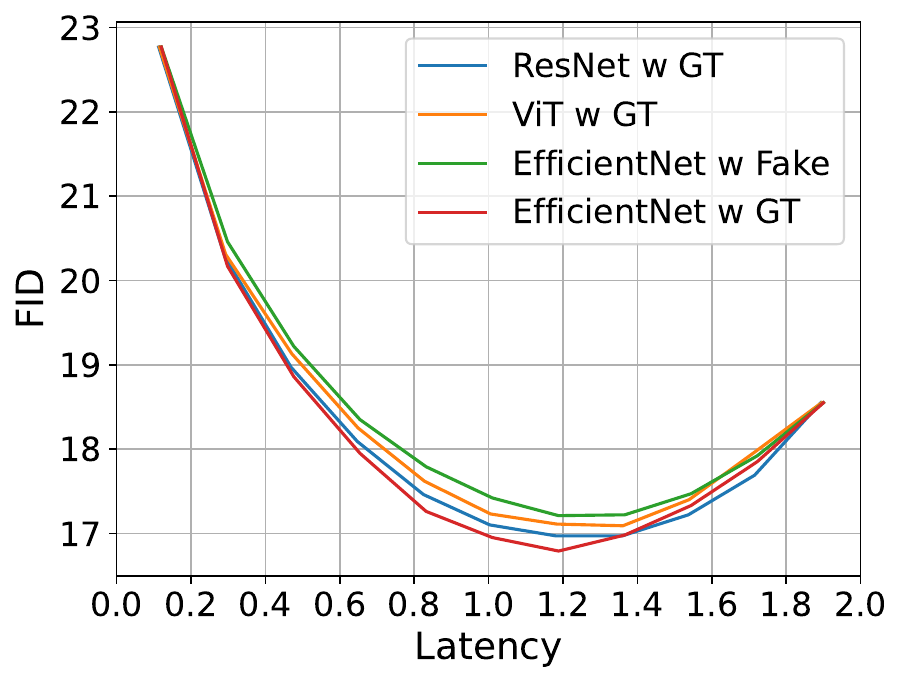}}}%
    \subfloat[\centering SDXS \label{fig:discriminator-sdxs}]{{\includegraphics[width=.49\linewidth]{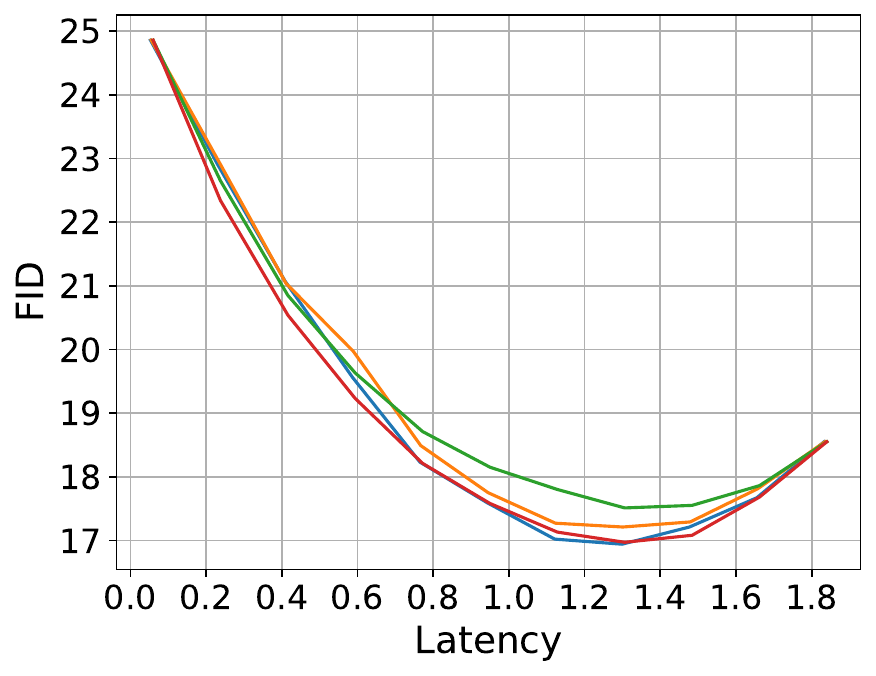}}}%
    % \vspace{-0.5em}
    \caption{Discriminator comparison. EfficientNet trained with ground truth images achieves the lowest FID given latency requirements, outperforming all other approaches across both cascades.}
    \label{fig:discriminator-choice}
    % \vspace*{-6mm}
\end{figure}

\begin{figure}[t]
  \centering
    \includegraphics[width=0.75\linewidth]{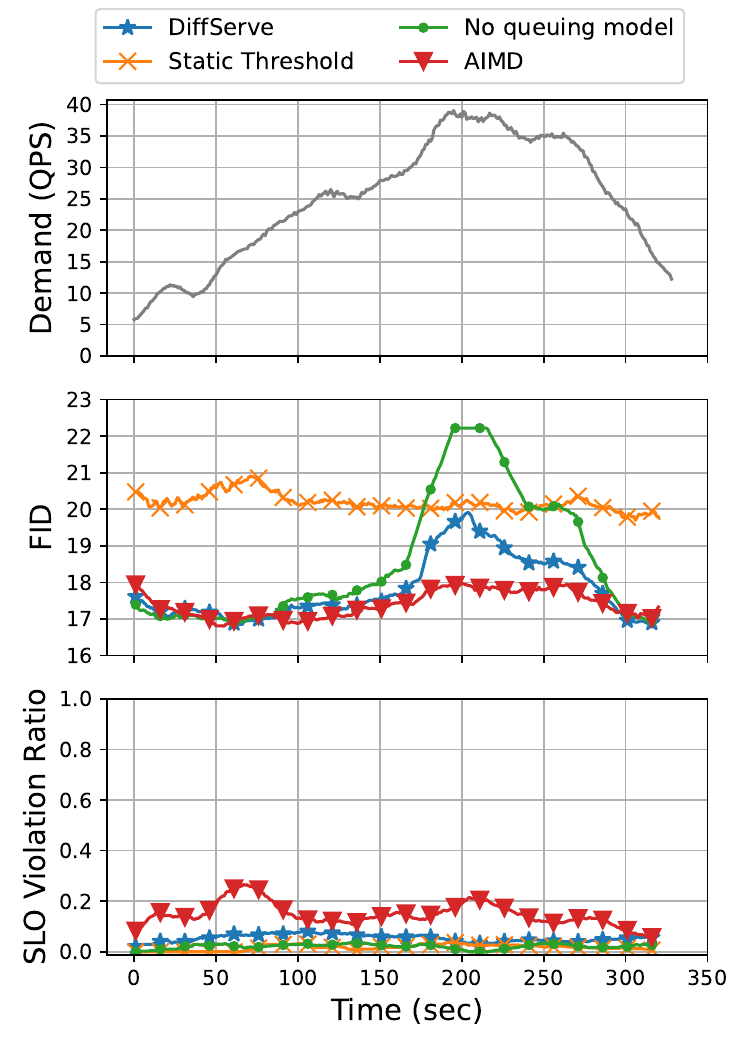}
    % \vspace{-1.3em}
  \caption{Performance of the resource allocation algorithm in \pjn{} and alternative approaches. \pjn{} reduces SLO violations by 20\% and improves quality by up to 19\%.}
  \vspace{-0.5em}
  \label{fig:ablation}
\end{figure}

We have shown in Figure~\ref{fig:model-variants} that using a discriminator surpasses other methods, such as PickScore, ClipScore, and Random selection, for diffusion model cascading. To further illustrate the choice of discriminator design, we examine the effectiveness of using \textit{EfficientNet-V2} as the discriminator in our model cascading architecture by comparing it against several variants of our approach. The variants include: (1) \textbf{ResNet w GT}: using ResNet-34~\cite{he2015deepresiduallearningimage} trained with ground truth images as ``real'' samples, (2) \textbf{ViT w GT}: using Vision Transformer~\cite{dosovitskiy2021imageworth16x16words} (i.e., ViT\_b\_16) trained with ground truth images as ``real'' samples, (3) \textbf{EfficientNet w Fake}: using EfficientNet-V2 trained with images generated by the heavymodel as ``real'' samples, and (4) \textbf{EfficientNet w GT}: using EfficientNet-v2 trained with ground truth images as ``real'' samples, which is the final configuration applied in our paper. We evaluate all the variants across two cascade configurations, where the lightweight models are SD-Turbo and SDXS, respectively, and the heavyweight model is SDV1.5 for both. The ground truth images are taken from the MS-COCO dataset.

Our results reported in Figure~\ref{fig:discriminator-choice} reflect that \textbf{EfficientNet w GT} consistently has the highest response quality (i.e., lowest FID scores) given a latency requirement among all the variants of approach, which means it outperforms all the other baselines. The latency of EfficientNet, ResNet, and ViT on an A100 GPU is 10ms, 2ms, and 5ms, respectively, which are negligible compared to the execution time of the diffusion models which are in the order of seconds. EfficientNet's architectural efficiency likely contributes to its superior performance, allowing it to capture complex quality features in images more effectively than the other models. Additionally, using ground truth images as ``real'' samples for training provides a more robust discriminator that aligns well with human-perceived quality, outperforming the configuration trained with generated images as ``real'' samples. These findings validate our choice of EfficientNet with ground truth images as the optimal configuration for the discriminator in \pjn{}, leading to improved accuracy in confidence estimation and overall performance within the cascading architecture.

\begin{figure}[t]
  \centering
    \includegraphics[width=0.9\linewidth]{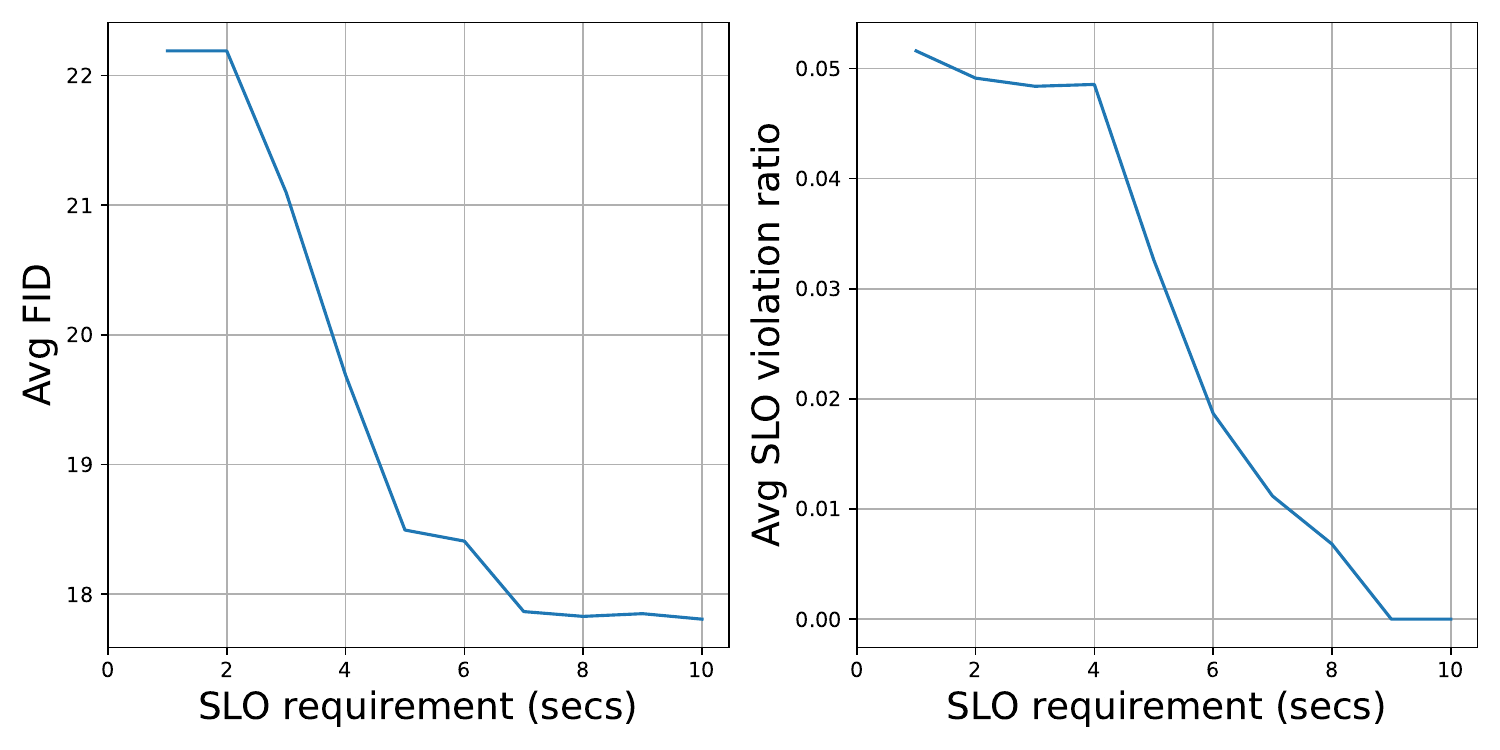}
    % \vspace{-1em}
  \caption{Effect of SLO on performance for Cascade 1. \pjn{} guarantees low SLO violations and high quality over a broad range of SLO values.}
  % \vspace{-1em}
  \label{fig:slo}
\end{figure}

\subsection{Evaluation of Resource Allocation}
\label{sec:eval_allocation}
% \TODO{Variants to compare with on dynamic trace (aggregated): 
% (1) not co-optimizing confidence threshold with the rest of parameters (essentially the same as query-aware static) 
% (2) instead of using a runtime-based queueing model but the naive version $L = 2 * L_{exe}$, 
% (3) AIMD 
% (4) Our approach}

% \begin{figure}[t]
%   \centering
%     \includegraphics[width=0.75\linewidth]{figures/ablation.pdf}
%     % \vspace{-1.3em}
%   \caption{Performance of the resource allocation algorithm in \pjn{} and alternative approaches. \pjn{} reduces SLO violations by 20\% and improves quality by up to 19\%.}
%   % \vspace{-1.5em}
%   \label{fig:ablation}
% \end{figure}

We now show an ablation study of our resource allocation in Figure~\ref{fig:ablation} to understand the effect of various parts of our optimization such as threshold tuning, dynamic batch size selection, and our queuing model, against the following:

% \begin{itemize}
\textbf{Static threshold.} We fix the threshold to remain static in \pjn{} throughout the experiment while letting the optimization tune server allocation and batch size. Note that this approach is different from \pjn{}-Static where all the optimization parameters are fixed and provisioned for the peak. As the threshold is fixed, this approach cannot adapt to demand changes, and thus loses out on the quality improvement offered by \pjn{} (up to 19\%) during off-peak by sending more queries to the heavyweight model.

\textbf{AIMD batching.} Instead of letting our optimization tune the batch size, we use a heuristic used by prior work such as Clipper~\cite{crankshaw2017clipper}: additive-increase multiplicative decrease. This heuristic decreases the batch size by a multiplicative factor upon experiencing an SLO timeout and increases it additively otherwise. As this approach is reactive based on SLO violations as signal, it experiences a significantly higher SLO violation ratio throughout the experiment. \pjn{} proactively sets the batch size based on the system demand and available resources, and thus experiences significantly lower SLO violations (up to 20\%).

\textbf{No queuing model.} Since \pjn{} uses a queuing model to estimate the queuing delays, we replace the theoretical model with a heuristic used by prior work such as Proteus to estimate queuing delays by assuming it to be twice the execution delay. This approach works well when query load is low and a query can always be executed in the next batch after it arrives. However, during off-peak times, it experiences significantly lower quality (up to 12\%) due to under-estimation of queuing delay. \pjn{} avoids this problem by estimating queuing delay based on real-time queue length and query arrival information.
% \end{itemize}

\textbf{Overhead of MILP Solver.} We measure the average runtime of the MILP solver to be $\sim$10 milliseconds. This incurs a minimal overhead on the system for periodically changing resource allocation in real-time. Moreover, since the MILP solver does not lie on the critical path of query serving, this overhead does not affect individual queries.
% \TODO{1-2 sentences of MILP scalability: is the no. of constraints linear in terms of no. of servers, other variables, etc}

% \begin{figure}[t]
%   \centering
%     \includegraphics[width=0.9\linewidth]{figures/slo.pdf}
%     % \vspace{-1em}
%   \caption{Effect of SLO on performance for Cascade 1. \pjn{} guarantees low SLO violations and high quality over a broad range of SLO values.}
%   % \vspace{-1em}
%   \label{fig:slo}
% \end{figure}

% \vspace{-1em}
\subsection{Sensitivity to SLO}
\label{sec:sensitivity}

Figure~\ref{fig:slo} explores the effect of different SLO values on the accuracy and SLO violation ratios. We note that \pjn{} consistently guarantees low SLO violations and high quality across a broad range of SLO values.

\section{Discussion}
\mypar{Scalability of \pjn{}.} 
\pjn{} is scalable and can be generalized to various situations. 
For longer pipelines, \pjn{} can be extended by applying a discriminator after each model, with adjustments to the MILP formulation to include multiple confidence thresholds as optimization variables. 
For higher-resolution image generation, the overhead of the discriminator remains negligible compared to the overall pipeline as the computational complexities of the discriminator and the diffusion models grow with the image resolution.
For other models and tasks (e.g., LLMs), the basic technique of model cascading is still applicable with specific quality metrics like BARTScore~\cite{NEURIPS2021_e4d2b6e6} used in the discriminator.
For deploying \pjn{} in heterogeneous GPU clusters, a slightly more complex MILP formulation would be required to account for different server classes and model runtimes on each class. Although this adjustment would increase the runtime complexity of the MILP, there is no fundamental limitation that prevents its implementation.

\mypar{Reuse Opportunities.}
A potential optimization to improve \pjn{} is the \textit{reuse} of intermediate outputs from lightweight models during the execution of heavyweight models. This approach compensates for the overhead of both difficulty estimation and the initial execution of lightweight models, as the heavyweight model can directly build upon the results from the lightweight model instead of starting from scratch. 
While reuse opportunities exist in cascaded inference, they introduce additional complexity in selecting compatible lightweight and heavyweight models, as reuse may even potentially negatively impact the image generation quality of the heavyweight model. For instance, in our experiments, with 50 denoising steps, reusing images from SD-Turbo with SDv1.5 showed no significant change in FID, but reusing those from SDXS increased FID from 18.55 to 19.75 on the MS-COCO dataset, implying worse image quality. While reuse is feasible, ensuring compatibility between models is critical to maintaining output quality.

\mypar{Design of Predictive Router.}
While \pjn{} relies on a discriminator to assess the quality of generated images for routing decisions, an alternative approach is to use the query itself to make routing decisions before executing any diffusion models. However, predicting image generation quality solely from text inputs is challenging, as image quality is highly dependent on the specific diffusion models used, making it non-trivial to accurately estimate outcomes based on the query alone. Thus, it remains an open question whether a query-based routing strategy would yield better performance in image-generation pipelines.
\section{Related Work}
% \vspace{-1em}
\mypar{Model Serving.}
Existing model serving systems generally fall into two categories, neither of which addresses the resource management challenges with query-aware model scaling. 
The first category requires users to specify the models for inference, leaving the system to handle resource management alone. This category includes production systems like SageMaker~\cite{sagemaker}, Triton~\cite{triton}, TensorFlow Serving~\cite{tensorflow_serving}, and TorchServe~\cite{pytorch_serve}, as well as academic prototypes such as Nexus~\cite{shen2019nexus}, BATCH~\cite{ali2020batch}, and Clockwork~\cite{gujarati2020serving}. 
% More recent systems optimized for Generative AI, like those in~\cite{agarwal2024approximate, li2023alpaserve, kwon2023efficient, chen2024punica, sheng2023s}, also fall under this category. Some other research efforts in this area focus on efficient GPU sharing across tasks~\cite{choi2022serving, jain2018dynamic, strati2024orion, ng2023paella, yu2019salus}. 
These systems are limited in that they do not automatically manage model variants for the same task, providing only a narrow scope of services.
The second category improves upon the first by automatically managing both resources and model variants. Works such as Clipper~\cite{crankshaw2017clipper}, Rafiki~\cite{wang2018rafiki}, and Cocktail~\cite{gunasekaran2022cocktail} use model ensembles to enhance response quality. INFaaS~\cite{romero2021infaas}, Model Switching~\cite{zhang2020model}, and Sommelier~\cite{guo2022sommelier} dynamically select model variants based on system load to process queries. 
% SuperServe~\cite{khare2023superserve} builds weight-shared supernetworks to adapt to query demand variations to avoid the overhead of storing multiple independent model variants. 
Proteus~\cite{ahmad2024proteus} is the first to formalize and address the resource management problem when serving model variants with different accuracy-efficiency tradeoffs. However, these systems route queries based primarily on system workload, overlooking the optimization potential inherent in varying query difficulties.   

\textbf{Cascaded Inference.} 
A cascading architecture for model inference has been explored across various machine learning domains in recent years. 
CascadeBERT~\cite{li2021cascadebertacceleratinginferencepretrained} and Tabi~\cite{10.1145/3552326.3587438} speed up inference by cascading progressively larger pre-trained language models, using calibrated confidence scores to decide whether to return results or re-route inputs to complex models.
\cite{8099719} introduces a cascade of CNN models with adaptive decision-making for efficient video classification.
WILLUMP~\cite{kraft2020willumpstatisticallyawareendtoendoptimizer} cascades feature computation by classifying simple inputs with inexpensive key features while routing complex ones to a more powerful model. 
For generative language models, \cite{gupta2024languagemodelcascadestokenlevel} leverages token-level uncertainty for deferral rules in LM cascades, and FrugalGPT~\cite{chen2023frugalgptuselargelanguage} proposes a flexible LLM cascade that learns optimal combinations of models to use for different queries.
Nonetheless, unlike discriminative inference and language generation tasks, diffusion models lack clear intermediate decision points to evaluate and cascade based on partial outputs or confidence, making these methods incompatible with text-to-image generation pipelines.

\textbf{Pipeline Serving.}
Works that serve inference pipelines are relevant as model cascades are a type of inference pipeline. 
InferLine~\cite{crankshaw2020inferline} minimizes the cost of inference serving by scaling hardware in response to changes in demand. 
A lot of work has been specifically related to video analytics pipelines such as VideoStorm~\cite{zhang2017videostorm}, Scrooge~\cite{hu2021scrooge}, Llama~\cite{llama2021romero}, and Nexus \cite{shen2019nexus}. 
% VideoStorm~\cite{zhang2017videostorm} was the first work to explore the latency-accuracy trade-off for the provisioning of resources for video analytics applications that use DNN. 
% Scrooge~\cite{hu2021scrooge} minimizes hardware costs for cloud-hosted DL-based media applications that process video and audio streams. 
% Llama~\cite{llama2021romero} is a serverless framework for auto-tuning video analytics pipelines. Nexus \cite{shen2019nexus} is another framework for serving video analytics pipelines on GPU clusters. 
% While these works aim to optimize resource allocation and performance in inference pipelines, 
\pjn{} differs in its focus on cascading models within the pipeline, using a confidence-based decision process to dynamically switch between lightweight and heavyweight models, balancing both computational efficiency and response quality.
\vspace{-1em}
\section{Conclusion}
\label{sec:conclusion}

This work proposed \pjn{}, an innovative system that optimizes the efficiency of text-to-image diffusion model serving by leveraging query-aware model scaling. \pjn{} employs a dynamic approach to model placement, implementing a cascading framework in which an ML-based discriminator routes queries based on their complexity, thereby preserving image generation quality.
By framing resource allocation as a mixed integer linear programming (MILP) problem, \pjn{} efficiently manages real-time query demand with optimized response quality. Our evaluations across various benchmarks indicate that \pjn{} enhances response quality by up to 24\% over existing systems while decreasing SLO violations by 19-70\%, demonstrating that query-aware scaling is both a robust and flexible solution for diffusion model serving.
% \appendix 

\section*{Acknowledgments}
This material is based upon work supported by the National Science Foundation under grants CNS-2106463, CNS-1901137, CNS-2312396, CNS-2338512, CNS-2224054, and DMS-2220211. Any opinions, findings, and conclusions or recommendations expressed in this material are those of the authors and do not necessarily reflect the views of the National Science Foundation. Part of this work is also supported by Adobe gift funding.

% \bibliography{bib}

\bibliographystyle{mlsys2025}

% LaTeX template for Artifact Evaluation V20190108
%
% Prepared by 
% * Grigori Fursin (cTuning foundation, France) 2014-2019
% * Bruce Childers (University of Pittsburgh, USA) 2014
%
% See example of this Artifact Appendix in
%  * SC'17 paper: https://dl.acm.org/citation.cfm?id=3126948
%  * CGO'17 paper: https://www.cl.cam.ac.uk/~sa614/papers/Software-Prefetching-CGO2017.pdf
%  * ACM ReQuEST-ASPLOS'18 paper: https://dl.acm.org/citation.cfm?doid=3229762.3229763
%
% (C)opyright 2014-2019
%
% CC BY 4.0 license
%

% \documentclass{sigplanconf}

% \usepackage{hyperref}
% \usepackage{xcolor}
% \newcommand{\TODO}[1]{\textcolor{blue}{\{\textit{TODO: #1}\}}}
% \newcommand{\highlight}[1]{{\color{blue} #1}}

% \begin{document}

% \special{papersize=8.5in,11in}

%%%%%%%%%%%%%%%%%%%%%%%%%%%%%%%%%%%%%%%%%%%%%%%%%%%%
% When adding this appendix to your paper, 
% please remove above part
%%%%%%%%%%%%%%%%%%%%%%%%%%%%%%%%%%%%%%%%%%%%%%%%%%%%
\clearpage
\setcounter{page}{1}

\appendix
\section{Artifact Appendix}

%%%%%%%%%%%%%%%%%%%%%%%%%%%%%%%%%%%%%%%%%%%%%%%%%%%%%%%%%%%%%%%%%%%%%
\subsection{Abstract}
This artifact describes the complete workflow to set up the cluster testbed experiments for \pjn{}. We introduce the hardware and software requirements of the cluster testbed. We then describe how to obtain the code, and then describe how to install the dependencies. Finally, we explain how to run the experiments and plot the results. 

\subsection{Artifact check-list (meta-information)}

{\small
\begin{itemize}
  \item {\bf Algorithm: } Optimization using a trained DNN model as a discriminator, combined with mixed integer linear programming (Gurobi is used).
  \item {\bf Data set: } Multiple text files included in the repository.
  \item {\bf Run-time environment: } Linux Ubuntu
  \item {\bf Hardware: } Multiple powerful GPUs (A100-40G/80G, L40s, etc.) for execution of diffusion model, or multiple Nvidia GPUs (1080Ti, V100, etc.) for simulated execution. 
  \item {\bf Metrics: } Confidence threshold, FID score, SLO violation ratio.
  \item {\bf Output: } Log files are output by the testbed, which are then used by the plotting scripts to generate graphs.
  \item {\bf Experiments: } End-to-end experiment of \pjn{} for three cascaded pipelines shown in the paper on dynamic traces.
  \item {\bf How much disk space required (approximately)?: } Approximately 15GB of disk space to download all diffusion models.
  \item {\bf How much time is needed to prepare workflow (approximately)?: } 30 minutes.
  \item {\bf How much time is needed to complete experiments (approximately)?: } 10 minutes for each cascaded pipeline.
  \item {\bf Publicly available?: } Yes. See Section~\ref{delivery} for details.
\end{itemize}}

%%%%%%%%%%%%%%%%%%%%%%%%%%%%%%%%%%%%%%%%%%%%%%%%%%%%%%%%%%%%%%%%%%%%%
\subsection{Description}

\subsubsection{How delivered} \label{delivery}
The cluster testbed code and workload trace can be accessed at \highlight{\url{https://github.com/qizhengyang98/DiffServe}}. It is also accessible at the following DOI: \highlight{\url{https://doi.org/10.5281/zenodo.14984970}}.

\subsubsection{Hardware dependencies}
The \pjn{} testbed requires a CPU server and multiple GPU servers.
{\small
\begin{itemize}
  \item {\bf CPU: } a CPU server with 10 cores, 16G RAM for the controller, load\_balancer, sink worker, and client processes.\\ 
  \item {\bf GPU: } multiple servers with powerful GPUs (e.g., A100-40G/80G, L40s, etc) to execute diffusion models. The number of servers depends on how many workers are used in the experiments. At a minimum, 4 servers are needed. In the paper, we used 16 servers with A100-80G for 16 workers. 
\end{itemize}}
Alternatively, we provide simulated execution in the artifact, which simulates the execution of diffusion models. In this case, any Nvidia GPUs can be used, and multiple workers can be created on a single GPU server by using \textit{tmux} such that the experiments can be done with fewer GPUs.

\subsubsection{Software dependencies}
{\small
\begin{itemize}
  \item Linux OS, conda environment with Python=3.8, and several Python packages listed in the \textit{requirements.txt}.
  \item Gurobi optimization software.
  \item A Gurobi license.
\end{itemize}}

\subsubsection{Datasets}
We have three cascaded pipelines: \textit{sdturbo\&sdv1.5}, \textit{sdxs\&sdv1.5}, and \textit{sdxl-lightning\&sdxl}. The name of the cascade indicates the lightweight and heavyweight models used. Each cascade uses a specific dataset and trace.

For query content, We use the MS-COCO 2017 dataset for Cascades 1 and 2, and DiffusionDB for Cascade 3. We select the first 5K text-image pairs from each dataset with text prompts serving as queries and the respective images used to calculate FID scores for evaluation. We provide text files in the artifact that are required to run the end-to-end experiments. 

For query arrivals, we use the Microsoft Azure Functions trace and scale it using shape-preserving transformations to match the capacity of our system. All traces used in the paper are included in the artifact. The trace files follow the naming format \textit{trace\_\{A\}to\{B\}qps.txt}, where A and B represent the minimum and maximum query rates in the trace, respectively. For cascade 1 and 2, We used \textit{4to32qps}, while for cascade 3, we used \textit{1to8qps}, on a cluster of 16 GPUs.

%%%%%%%%%%%%%%%%%%%%%%%%%%%%%%%%%%%%%%%%%%%%%%%%%%%%%%%%%%%%%%%%%%%%%
\subsection{Installation}
To set up the environments, first clone the repository via the link in Section~\ref{delivery}. Then, go to the root folder and create a conda environment:
{\small
\begin{itemize}
  \item conda create -n diffserve python=3.8
  \item conda activate diffserve
  \item pip install -r requirements.txt
\end{itemize}}
We implement the MILP optimization of \pjn{} using Gurobi. Therefore, it is required to obtain a Gurobi license as follows:
{\small
\begin{itemize}
  \item Follow the instructions on the \href{https://www.gurobi.com/solutions/licensing/}{\highlight{official website}} to get a commercial or a free academic license for Gurobi.
  \item Once the license is obtained, Gurobi will provide a \textit{gurobi.lic} file.
  \item Place the license file under the path \textit{gurobi/gurobi.lic}.
\end{itemize}}
Then, download the pretrained discriminators and image datasets by running ``\textit{python prepare\_ds\_mod.py}'' under the conda environment. 
Note that this step can be skipped if you use simulated execution.

%%%%%%%%%%%%%%%%%%%%%%%%%%%%%%%%%%%%%%%%%%%%%%%%%%%%%%%%%%%%%%%%%%%%%
\subsection{Experiment workflow}
In this section, we provide instructions on how to execute experiments using the scripts in the artifact. In the following steps, \textit{\{Step Num.-R\}} means steps for doing real execution of diffusion models, while \textit{\{Step Num.-S\}} means steps for doing simulated execution.
{\small
\begin{itemize}
  \item {\bf Step 1-R} For preparation, open one terminal on each GPU server and 4 terminals on the CPU server. \\
  \item {\bf Step 2-R} In the terminals of the CPU server, run \textit{\{tmux new -s contr\}}, \textit{\{tmux new -s loadb\}}, \textit{\{tmux new -s sink\}}, \textit{\{tmux new -s client\}}, respectively. \\
  \item {\bf Step 3-R} Activate the conda environment in each terminal by running \textit{\{conda activate diffserve\}}.\\
  \item {\bf Step 4-R} To run experiments of cascade-1, in \textbf{tmux contr} terminal, run the script \textit{start\_controller.sh} which starts the controller process. Copy the IP address printed in the console and replace the original IP address after \textit{-cip} in \textit{start\_worker.sh} and \textit{start\_worker\_sink.sh}.\\
  \item {\bf Step 5-R} In \textbf{tmux loadb} terminal, run \textit{start\_load\_balancer.sh} which starts the load\_balancer process.\\
  \item {\bf Step 6-R} In \textbf{tmux sink} terminal, run \textit{start\_worker\_sink.sh} which starts the sink worker process.\\
  \item {\bf Step 7-R} In each terminal of GPU server, modify the number after \textit{-p} in \textit{start\_worker.sh}, then run the script. This number is the port number of each worker. Make sure the number you assign to each worker is unique, and is between [50051, 50066]. Note that if you execute the worker for the first time, diffusion models will be downloaded automatically to \textit{models} folder.\\
  \item {\bf Step 8-R} If all the processes have been set up successfully, there will be logs corresponding to each process under the folder \textit{logs}, and worker processes will report ``Worker is ready'' in the console. The logs include \textit{controller}, \textit{load\_balancer}, \textit{worker\_\{port number\}}, and \textit{model\_\{port number\}}.\\
  \item {\bf Step 9-R} Then in \textbf{tmux client}, run \textit{start\_client.sh} to start the client process, which keeps sending queries in 6 minutes. Modify the flag \textit{-trace} given the number of workers you allocate. Use \textit{1to8qps}, \textit{2to16qps}, \textit{2\_5to20qps}, \textit{3to24qps}, \textit{4to32qps} if you have $\sim$4, 8, 10, 12, 16 workers, respectively.\\
  \item {\bf Step 10-R} The Client process will report \textit{``Trace ended''} when it stops sending queries. Then stop all the processes.\\
  \item {\bf Step 11-R} Under folder \textit{logs}, there will be three \textit{csv} files which contains the end-to-end experiment results. To generate graphs, go to folder \textit{plotting}, and run the script \textit{run\_plot\_results.sh}. \\
  \item {\bf Step 12-R} Be sure to remove all the log files before starting a new experiment. \\
\end{itemize}}

For simulated execution,
{\small
\begin{itemize}
  \item {\bf Step 1-S} For preparation, open one or multiple terminals on each GPU server and 4 terminals on the CPU server. The total number of terminals on GPU servers should be equal to the total number of workers you want to allocate.\\
  \item {\bf Step 2.1-S} In the terminals of the CPU server, run \textit{\{tmux new -s contr\}}, \textit{\{tmux new -s loadb\}}, \textit{\{tmux new -s sink\}}, \textit{\{tmux new -s client\}}, respectively. \\
  \item {\bf Step 2.2-S} In each terminal of the GPU servers, run \textit{\{tmux new -s workerX\}} respectively, where \textit{X} is a number or character. Make sure \textit{X} is unique to each terminal on a single server. \\
  \item {\bf Step 2.3-S} In \textit{start\_worker.sh}, add a flag \textit{-{}-do\_simulate} at the end of the python command.\\
  \item {\bf Step (3-12)-S} Steps 3-12 are the same as Step 3-R to Step 12-R explained above. \\
\end{itemize}}
The above steps describe the end-to-end experiment flow of cascade 1. To run experiments for Cascades 2 and 3, simply replace the flag \textit{-c sdturbo} with \textit{-c sdxs} and \textit{-c sdxlltn} in all shell scripts in the root folder, then follow the same steps. For Cascade 3, it is recommended to use simulated execution when the number of GPUs is less than 16, as the controller may struggle to find a solution due to insufficient available workers.

%%%%%%%%%%%%%%%%%%%%%%%%%%%%%%%%%%%%%%%%%%%%%%%%%%%%%%%%%%%%%%%%%%%%%
\subsection{Evaluation and expected result}
The testbed produces log files in the \textit{logs} folder. The log files contain snapshots of the system at regular intervals, including the resource allocation, user demands, system capacity, queries served/dropped/late, and confidence thresholds set given the demand changes.

To generate the graphs from the logs, the Python script \textit{plotting/plot\_results.py} can be used to plot the following: confidence threshold over time, SLO violation ratio over time, and FID score over time, which should be similar to the one in Figure 5. The script also prints the average SLO violation ratio and the average FID score, which should be similar to those in Figure 6. The results can vary slightly given different hardware and trace files in use. To generate the graphs, simply modify the flag \textit{-{}-cascade} with [sdturbo, sdxs, sdxlltn] for different cascades, then run the script \textit{run\_plot\_results.sh}. 

%%%%%%%%%%%%%%%%%%%%%%%%%%%%%%%%%%%%%%%%%%%%%%%%%%%%%%%%%%%%%%%%%%%%%

% \end{document}

\end{document}